\documentclass{aa}

\usepackage{graphicx}
\usepackage[section]{placeins}
\usepackage{sidecap}
\usepackage[english]{babel}
\usepackage{multirow}
\usepackage{color}
\usepackage{url}
\usepackage{amsmath}
\usepackage{amssymb}

\usepackage{natbib}
\bibpunct{(}{)}{;}{a}{}{,}

\def\be{\begin{equation}}
\def\ee{\end{equation}}
\def\bea{\begin{eqnarray}}
\def\eea{\end{eqnarray}}
\def\ba#1\ea{\begin{align}#1\end{align}}

\def\dd{\mathrm{d}}
\def\xx{\mathbf{x}}
\def\kk{\mathbf{k}}
\def\hn{\mathbf{\hat{n}}}

\def\mr{\mathrm}
\def\lbra{\left\langle}
\def\rbra{\right\rangle}
\def\Cov{\mr{Cov}}
\def\Msun{M_{\odot}}
\def\Ncl{\ensuremath{N_\mr{cl}}}

\def\nbargal{\ensuremath{\overline{n}_\mathrm{gal}}}

\newcommand{\threeJz}[3]{\begin{pmatrix}
#1 & #2 & #3 \\ 0 & 0 & 0
\end{pmatrix}}

\begin{document}

\title{Super-sample covariance approximations and partial sky coverage}
\titlerunning{SSC approximations and partial sky}

\author{Fabien Lacasa\inst{\ref{inst1}}\thanks{fabien.lacasa@unige.ch}
\and  Marcos Lima\inst{\ref{inst2}}
\and  Michel Aguena\inst{\ref{inst2}}
}
\institute{
D\'{e}partement de Physique Th\'{e}orique, Universit\'{e} de Gen\`{e}ve, 24 quai Ernest Ansermet, CH-1211 Geneva, Switzerland\label{inst1} 
\and Departamento de F\'{\i}sica Matem\'atica,
Instituto de F\'{\i}sica, Universidade de S\~ao Paulo, CP 66318, CEP 05314-970, S\~ao Paulo-SP, Brazil\label{inst2} 
}

\date{\today}

\abstract
{
Super-sample covariance (SSC) is the dominant source of statistical error on large scale structure (LSS) observables for both current and future galaxy surveys.
In this work, we concentrate on the SSC of cluster counts, also known as sample variance, which is particularly useful for the self-calibration of the cluster observable-mass relation; our approach can similarly be applied to other observables, such as galaxy clustering and lensing shear.
We first examined the accuracy of two analytical approximations proposed in the literature for the flat sky limit, finding that they are accurate at the 15\% and 30-35\% level, respectively, for covariances of counts in the same redshift bin.
We then developed a harmonic expansion formalism that allows for the prediction of SSC in an arbitrary survey mask geometry, such as large sky areas of current and future surveys. We show analytically and numerically that this formalism recovers the full sky and flat sky limits present in the literature. We then present an efficient numerical implementation of the formalism, which allows fast and easy runs of covariance predictions when the survey mask is modified. We applied our method to a mask that is broadly similar to the Dark Energy Survey footprint, finding a non-negligible negative cross-z covariance, i.e. redshift bins are anti-correlated. We also examined the case of data removal from holes due to, for example bright stars, quality cuts, or systematic removals, and find that this does not have noticeable effects on the structure of the SSC matrix, only rescaling its amplitude by the effective survey area.
These advances enable analytical covariances of LSS observables to be computed for current and future galaxy surveys, which cover large areas of the sky where the flat sky approximation fails.
}

\maketitle

%%%%%%%%%%%%%%%%%%%%%%%%%%%%%%%%%%%%%%%%%%%%%%%%%%%%%%%%%%%%%%%%%%%

\section{Introduction}\label{Sect:intro}

%Large scale structure and covariances
The large scale structure (LSS) of the universe results from the growth of local density perturbations induced by gravitational collapse within an expanding background \cite[e.g.][]{Peebles1980}. The multi-point correlations and associated spectra that characterise this structure can be measured in 
real data and combined with theory predictions to constrain cosmological models, including gravity theories and relative amounts of dark matter and dark energy \citep{Astier2006,Percival2010,Riess2009,Hildebrandt2016,Beutler2017,Hinton2017,DES2017}. However, likelihood methods of parameter inference also require reliable estimates of the covariances associated with structure observables. Estimates of the covariances can be obtained from the actual data via methods such as jackknife and bootstrap, from simulations and Monte Carlo realisations, and from theoretical predictions that account for the known observational effects \citep{Dodelson2013, Giannantonio2016,Crocce2016,OConnell2016,Singh2016,Shirasaki2016,Blot2016,Escoffier2016,Pearson2016, Camacho_etal_inprep}. In this work we focus on the latter approach.

%Cluster counts and cluster covariance
Within the so-called halo model paradigm \citep{Cooray2002}, the LSS can be characterised by the statistical properties of the universe building blocks, i.e. dark matter haloes. These haloes are characterised by their abundance, bias, and profiles, all of which can be studied from dark matter N-body simulations and also from high-quality data sets. As galaxy clusters develop within dark matter haloes, they trace the highest density peaks. Their number counts and covariances are very sensitive probes of structure growth and expansion of the Universe \citep{Lima2004, Lima2005, Lima2007,Schmidt2009, Aguena2016}. A number of past and current surveys have detected clusters in multiple wavelengths with hopes to use such detections for cosmological purposes \citep{Miller2005, Koester2007, Soares-Santos2011,Dietrich2014, Rykkoff2014,Bleem2015,Ade2016,  Rykoff2016,Bayliss2016}.

%Super-sample covariance
As we consider scales close to the survey maximum size, the number of modes available decreases significantly, which represents an intrinsic source of uncertainty for the inferred structure properties. The cosmic variance quantifies these uncertainties and may have contributions from scales much larger than the survey itself, in which case we refer to it as super-sample covariance (SSC), whose effects have been studied recently in multiple contexts \citep[e.g.][]{Takada2013, Li2014, Li2014b, Takahashi2014, Li2016, Shirasaki2016, Hu2016}. SSC has been shown to be particularly important for probe combinations as it couples observables together \citep{TakadaBridle2007, Takada2014, Krause2016, Lacasa2016}. Within a local patch, large scale modes change the effective average density, which can differ significantly and unpredictably from the true background density, affecting the inferred correlations. In fact, SSC may be the dominant source of errors in Jackknife covariance estimations \citep{Shirasaki2016}.

%Partial sky coverage and masks
Finally, the survey geometry or footprint and its selection function, which is characterised by, for example masks and depth maps, also affect the estimation of correlations and covariances
\citep{Takahashi2014}. 
Not only the geometry and selection must be known to good precision, but their properties must be properly propagated into the measured and predicted correlations and covariances. 
Some predictions can only be directly made under certain approximations, for example for full sky calculations, while for the more realistic case of partial sky coverage, further complicating assumptions need to be made. As we attempt to extract maximum information from observations, we may end up with a complicated survey mask containing holes, for example owing to saturated pixels, asteroids, contamination from stars, or simply pixels that do not satisfy a depth criterion. All these effects must be accounted for in a proper cosmological analysis. 

%Goals of the paper
In this article, we study the effect of partial sky coverage and  arbitrary masks in the estimation of the SSC of cluster counts. We propose a method to estimate this covariance  efficiently for an arbitrary mask, and compare this approach to a number of approximated calculations in the literature. We show that our general calculation reduces to the approximated computations in the appropriate limits, but differs from such computations in general. Although beyond the scope of this work, we expect our theoretical estimation can be compared to other methods of estimating the covariance, which also attempt to account for the survey geometry and mask effects such as with simulations. One of the advantages of our method, however, is that it allows for the covariance to be computed as a function of cosmological and nuisance parameters at each step of a Monte Carlo Markov chain, allowing, for example for self-calibration of cluster observable-mass distribution in general cosmological analyses of cluster samples \citep{Lima2005}. It is also unbiased, contrary to internal covariance estimators \citep{Lacasa2017}. Finally, it is numerically much cheaper than running thousands of N-body simulations, the latter needing to be much larger than the survey to capture super-survey modes efficiently.

%article summary
This article is organised as follows. In Sect.~\ref{Sect:SSC}, we introduce the cluster counts covariance and the formalism for an exact SSC computation, and we then consider the case of full sky and flat sky (small angles) limit, providing some comparisons for the covariance kernels between those cases. In Sect.~\ref{Sect:comp-approxs} we consider the flat sky limit under various approximations that have been proposed in 
the literature for computing the SCC and compare the results to the exact computation. In Sect.~\ref{Sect:psky-derivation}, we propose a method to numerically compute SSC in the case of partial sky coverage with an arbitrary survey mask. In Sect.~\ref{Sect:results} we present our results, applying the method to the case of a geometry similar to that of the Dark Energy Survey (DES), discussing the mask effects and recovering the flat sky limit. Finally, in Sect.~\ref{Sect:conclusion} we present our conclusions and perpectives.

%Extra facts
In all numerical computations, we take a cosmology for a flat $\Lambda$CDM universe with parameter values  $h=0.67$, $\Omega_b h^2=0.022$, $\Omega_c h^2=0.12$, $w=-1$, $n_S=0.96$ , $\sigma_8=0.83$.
Cluster counts are computed in two bins of redshift in the range $z\in[0.4,0.6]$ with $\Delta z=0.1$, and four bins of mass in the range $\log [M/(h^{-1}\Msun)] \in[14,16]$ 
with $\Delta\log [M/(h^{-1}\Msun)]=0.5$. The halo mass function is from a fit to simulations from \cite{Tinker2008}, the halo bias is from \cite{Tinker2010}, and the linear matter power spectrum is from the transfer function by \cite{Eisenstein1998}.
We set the following notational conventions: $r(z)$ is the comoving distance, that is $dr=c\, dz/H(z)$ with $H(z)$ the Hubble parameter, and  $\dd V = r^2 dr$ is the comoving volume element per steradian. We use short cuts such as $\dd X_{12}= \dd X_1 \, \dd X_2$ and $P_m(k|z_{12})=P_m(k|z_1,z_2)=G(z_1)G(z_2)P_m(k)$ with $G(z)$ the linear growth function. Often the limits of redshift or mass integrals are implicitly those of the redshift or mass bin considered.

%%%%%%%%%%%%%%%%%%%%%%%%%%%%%%%%%%%%%%%%%%%%%%%%%%%%%%%%%%%%%%%%%%%%
\section{Covariance of large scale structure observables and SSC}\label{Sect:CovLSS-and-SSC}

%%%%%%%%%%%%%%%%%%%%
\subsection{Covariance of large scale structure observables}\label{Sect:CovLSS}

Large scale structure observables have a covariance that can be composed of many different terms \citep[e.g.][]{Lacasa2016,Lacasa2017b}. The particular case of interest to this article is that of cluster counts, which can be viewed as a halo 1-point correlation function summed over the survey area. 

We address the covariance as given by a decomposition in two specific terms: 1-halo and 2-halo
\be
\Cov(N_\mr{cl}(i_M,i_z),N_\mr{cl}(j_M,j_z)) = \Cov^\mr{1h} + \Cov^\mr{2h}\,.
\ee
These terms are denoted respectively as shot-noise variance and sample (co)variance in the literature \citep[e.g.][]{Hu2003}.

The 1-halo or shot-noise variance term is given by
,\ba\label{Eq:shotnoise}
\Cov^\mr{shot}(N_\mr{cl}(i_M,i_z),N_\mr{cl}(j_M,j_z)) =  \frac{N_\mr{cl}(i_M,i_z)}{\Omega_S} \; \delta_{i_M,j_M} \; \delta_{i_z,j_z}
\ea
where $\Omega_S$ is the solid angle covered by the survey. In the following discussion, this term is accounted for when showing the total counts covariance. Its dependence on the survey footprint is a trivial scaling by $\Omega_S$ \footnote{We note that the factor $\Omega_S$ appears because in our convention $N_\mr{cl}$ is a number of objects per steradian. In data analysis it is sometimes more convenient to use the absolute number of objects $\tilde{N}_\mr{cl}=N_\mr{cl} \times \Omega_S$, for which we have the simpler formula $\mr{Var}(\tilde{N}_\mr{cl}(i_M,i_z))=\tilde{N}_\mr{cl}(i_M,i_z)$} and does not require any particular formalism to be predicted for a survey with large and/or complicated coverage.

We refer to the 2-halo term  as SSC, which is described in more detail in Sect. \ref{Sect:SSC}.

For observables other than cluster number counts \cite[for instance the galaxy angular power spectrum; see e.g.][]{Lacasa2016, Lacasa2017b} the covariance may contain many more terms, whose dependence on the survey geometry may be far from trivial. In this article, we restrict ourselves to studying the geometry dependence of SSC, as needed for cluster number counts analyses.

%%%%%%%%%%%%%%%%%%%%
\subsection{Super-sample covariance}\label{Sect:SSC}

Super-sample covariance is a source of uncertainties for LSS observables coming from modes of size larger than the survey. The effect of these large scale modes on structure comes from the fact that the effective or local matter density $\widetilde{\rho}_m(z)$ averaged within the survey can be different from the true background density $\bar{\rho}_{m}(z)$ averaged over the whole universe or an ensemble of universes written as
\be
\widetilde{\rho}_m(z)=\bar{\rho}_m(z)[1+\delta_b(z)]\,,
\ee
where  $\delta_b(z)$ is a background density perturbation induced by large scale modes. 
Equivalently, defining the density contrasts
\be
\delta_m(\xx)=\frac{\rho(\xx)-\bar{\rho}_m}{\bar{\rho}_m},
\ee
the ensemble average is $\langle \delta_m(\xx)\rangle=0$ by definition, however the spatial average over the survey is $\langle\delta_m(\xx)\rangle_\mr{survey}=\delta_b$.

We split the survey window function $W(\xx)$ into its radial $W_r(z)$ and angular $W(\hn)$ pieces, i.e.  $W(\xx)=W_r(z)W(\hn)$. The radial part simply specifies the redshift binning and we do not explicitly state this in our description below anymore because it is effectively included with implicit bounds on redshift integrals. On the other hand, the angular window $W(\hn)$ is the main object whose effect on the covariances we want to consider. In principle $W(\hn)$ may depend on redshift as well. In fact, this is very often the case for surveys with significant depth variations across the sky. For simplicity, we keep the angular window only as a function of the angular vector, although the formalism can be easily generalised (see Appendix~\ref{App:maskz}).  

If the survey angular window $W(\hn)$ subtends a solid angle $\Omega_S$ in the sky, and we denote the position vector $\xx=r(z)\hn$, the background perturbation $\delta_b$ is  given by
\be\label{Eq:def-deltab}
\delta_b(z) = \frac{1}{\Omega_S} \int \dd^2\hn \ W(\hn) \ \delta_m(r(z)\hn,z)\,.
\ee

As a result, all LSS observables respond to such change of background density, becoming correlated. Explicitly, the SSC term of the cross-covariance between two observables $\mathcal{O}_1$ and $\mathcal{O}_2$ is given by \citep[e.g.][]{Lacasa2016}~
\be\label{Eq:SSC-general}
\Cov^\mr{SSC}(\mathcal{O}_1,\mathcal{O}_2) = \int \dd V_{12} \, \frac{\partial \mathfrak{o}_1}{\partial \delta_b} \, \frac{\partial \mathfrak{o}_2}{\partial \delta_b} \, \sigma^2(z_1,z_2)\,,
\ee
where $\mathfrak{o}_i$ is the density of observable $\mathcal{O}_i$ (per comoving volume $dV$),
$\partial \mathfrak{o}_i /\partial \delta_b$ 
is its reaction to the change in background $\delta_b$, and $\sigma^2(z_1,z_2)$ is the covariance of $\delta_b$~, i.e.
\ba
\nonumber \sigma^2(z_1,z_2) = \lbra \delta_b(z_1) \, \delta_b(z_2)\rbra 
&= \int \frac{\dd^3 \kk}{(2\pi)^3} \; \tilde{W}(\kk,z_1) \tilde{W}^*(\kk,z_2) \\ 
& \qquad \qquad \times P_m(k|z_{12})\,,
\ea
where $\tilde{W}$ is the Fourier transform of the survey window function.

In this article, we are mostly interested in the case of cluster number counts $N_\mr{cl}(i_M,i_z)$ per steradian within bins of mass $i_M$ and bins of redshift $i_z$ \footnote{Throughout this article $N_\mr{cl}$ is the number of clusters per steradian. For application to a survey, $N_\mr{cl}$ and its covariance may be rescaled by the observed area.}, 
\ba
N_\mr{cl}(i_M,i_z) = 
\int_{z\in \mr{bin}(i_z)} \dd V 
\int_{M\in \mr{bin}(i_M)} \dd M \, \frac{\dd n_h}{\dd M}(M,z) \,,
\ea
where $\dd n_h/\dd M$ is the halo mass function. 
The SSC of these counts is traditionally denoted simply as sample (co)variance in the cluster literature \citep{Hu2003}.
It is important to notice though that super-horizon modes affect even full sky surveys and must always be accounted for.
The SSC for clusters counts is given by \citep[e.g.][]{Lacasa2016}~
\ba\label{Eq:SSC_Ncl}
\Cov^\mr{SSC}(N_\mr{cl}(&i_M,i_z),N_\mr{cl}(j_M,j_z)) =  \text{} \nonumber \\ &\int \dd V_{12} \, \frac{\partial n_h}{\partial \delta_b}(i_M,z_1) \, \frac{\partial n_h}{\partial \delta_b}(i_M,z_2) \ \sigma^2(z_1,z_2)\,,
\ea
where the response of the halo number density $n_h=dN_\mr{cl}/dV$ to a change of background density is given by the number-weighted halo-bias averaged within the bin \citep[e.g.][]{Schmidt2013}~ as follows:
\be
\frac{\partial n_h}{\partial \delta_b}(i_M,z) = \int_{M\in \text{bin($i_M$)}} \dd M \, \frac{\dd n_h}{\dd M} \, b(M,z)\,,
\ee
where $b(M,z)$ is the (first order) halo bias.

Most of the elements for the computation of the covariance pose no particular numerical problem. A notable exception is $\sigma^2(z_1,z_2)$, as it depends on the survey geometry. Numerically tractable formulae are known only for a few cases with simple geometries.
First, for the case of full sky, we have \citep{Lacasa2016}~
\be\label{Eq:sigma2_fullsky}
\sigma^2_\mr{fullsky}(z_1,z_2) = \frac{1}{2\pi^2} \int k^2 \, \dd k \, j_0(k r_1) \, j_0(k r_2) \, P_m(k|z_{12}) \,,
\ee
where $r_i=r(z_i)$. 
Second, an expression for $\sigma^2(z_1,z_2)$ is also known in the flat sky limit \citep[e.g.][]{Hu2003, Lima2007, Aguena2016}~
\ba\label{Eq:sigma2_flatsky}
\sigma^2_\mr{flatsky}(z_1,z_2) = \frac{1}{2\pi^2} &\int k_\perp \, \dd k_\perp \, 4 \frac{J_1(k_\perp \theta_S r_1)}{k_\perp \theta_S r_1} \frac{J_1(k_\perp \theta_S r_2)}{k_\perp \theta_S r_2} \nonumber \\ \times &\int \dd k_\parallel \; \cos\left[k_\parallel (r_1-r_2)\right] \; P_m(k|z_{12}) \,,
\ea
for a cylindrical window function of radius $\theta_S$ delineating a survey solid angle $\Omega_S=2\pi(1-\cos\theta_S)\approx \pi\theta_S^2$, and the wave-vector $\kk=(k_\parallel, k_\perp)$ is split into its components parallel and perpendicular to the line of sight, where $k^2=k_\perp^2+k_\parallel^2$.

\begin{figure}[!th]
\begin{center}
\includegraphics[width=0.99\linewidth]{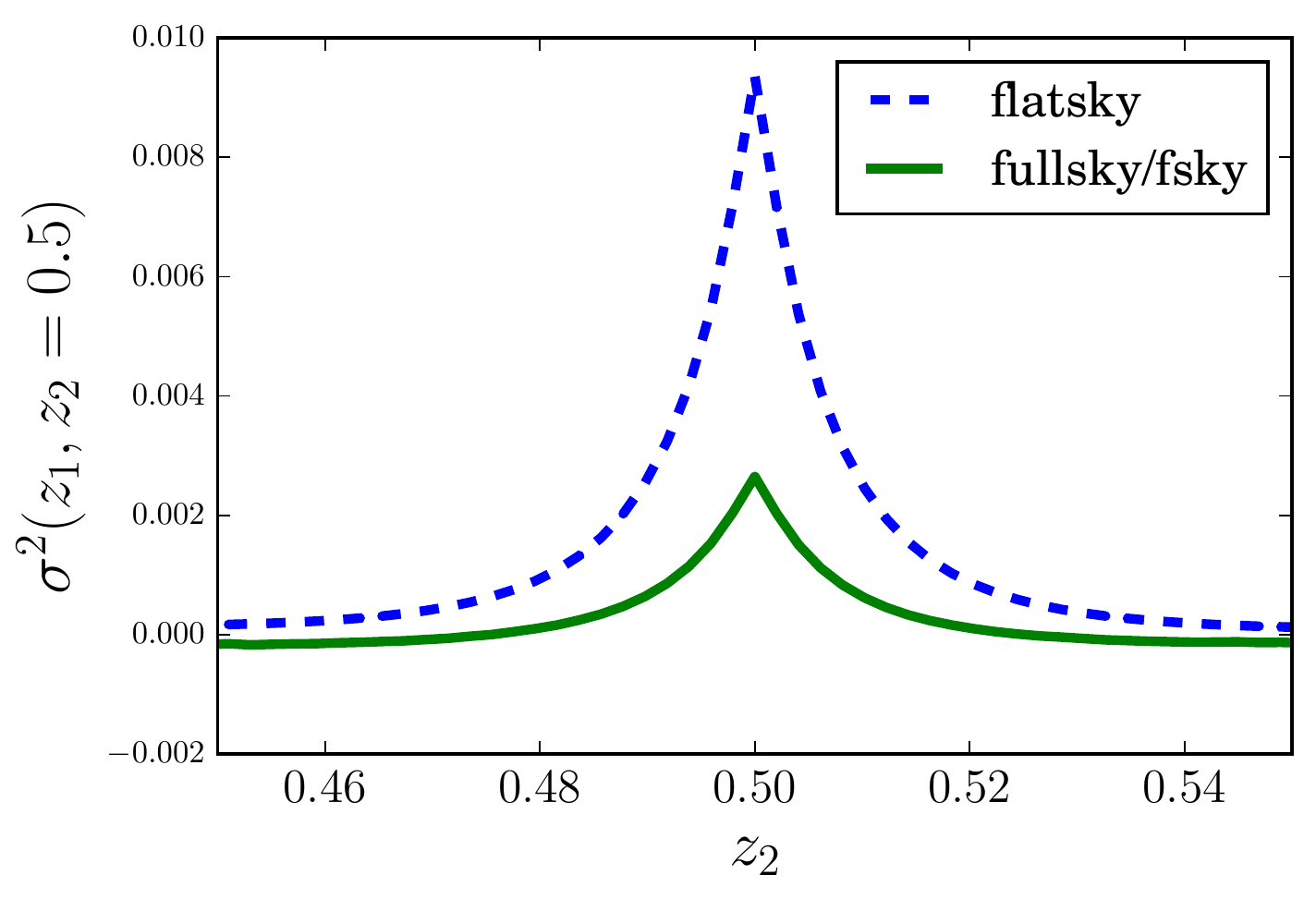}
\caption{Comparison of $\sigma^2(z_1,z_2)$ for $z_1=0.5$ in different cases. In blue the flat sky formula Eq.~(\ref{Eq:sigma2_flatsky}) for a survey angular radius $\theta_S=5$ deg. In green the full sky formula Eq.~(\ref{Eq:sigma2_fullsky}) rescaled by a factor $1/f_\mr{SKY}$.}
\label{Fig:sigma2-flatvsfull}
\end{center}
\end{figure}

Fig.~\ref{Fig:sigma2-flatvsfull} shows the comparison of $\sigma^2$ resulting from the two expressions above for an angular window of radius $\theta_S=5$ deg. The full sky covariance was rescaled by a factor $f_\mr{SKY}^{-1}$, where $f_\mr{SKY}=\Omega_S/4\pi$ and $\Omega_S \simeq \pi \theta_S^2$ for $\theta_S$ translated in radians.

We see that both covariances share the behaviour of peaking at $z_1=z_2$ and decreasing to zero as $|z_1-z_2|$ increases\footnote{By eye, both functions seem symmetric around $z_1=z_2$, but in detail this is not true.}. Regarding the amplitudes, we see that rescaling the full sky covariance by the usual $1/f_\mr{SKY}$ factor underpredicts the covariance by a factor $\sim 3.4$ in this case\footnote{An artificial factor $1/f_\mr{SKY}^2$ would not fare better, this time overpredicting the covariance.}.  Even when rescaling  the covariances to the same peak amplitude by
hand, we find that their shapes are broadly similar but differ in details. The full sky covariance is more strongly peaked at the centre, then decreases and crosses zero, reaching a negative minima of height $\sim 7\%$ of its peak, before slowly asymptoting zero. By contrast, the flat sky covariance is a bit broader in its positive part, but its negative minimum is only $\sim0.8\%$ of its peak, and it asymptotes faster to zero.  
This clearly indicates that SSC is non-trivially related to sky coverage, and a more general approach is required for its accurate computation.

%%%%%%%%%%%%%%%%%%%%%%%%%%%%%%%%%%%%%%%%%%%%%%%%%%%%%%%%%%%%%%%%%%%%
\section{Comparison of super-sample covariance approximations}\label{Sect:comp-approxs}

The full equation Eq.~(\ref{Eq:SSC_Ncl}) for SSC can be numerically expensive, as it requires a double redshift integral. Furthermore, for each pair of redshift, a double $(k_\perp,k_\parallel)$ integral is required (e.g. in the flat sky case). To tame down this burden, several approximations have been devised in the literature. 
Below we consider two approximations in the flat sky regime.

First, in Eq.~(\ref{Eq:SSC_Ncl}) we can assume that $\partial n_h/\partial \delta_b$ varies slowly with redshift within the bins, such that it can be approximated by its bin-averaged value and taken out of the integral. In fact this may be a good approximation for sufficiently narrow bins. In this case, Eq.~(\ref{Eq:SSC_Ncl}) takes the form \citep{Hu2003}
\ba 
\Cov^\mr{SSC}(N_\mr{cl}&(i_M,i_z),N_\mr{cl}(j_M,j_z)) \approx \nonumber \\ & N_\mr{cl}(i_M,i_z) \, b(i_M,i_z) \; N_\mr{cl}(j_M,j_z) \, b(j_M,j_z) \,
S_{i_z,j_z}\,, \label{Eq:cov-SSC-Sij}
\ea
where
\be
S_{i_z,j_z} =  \int \frac{\dd V_{12}}{V_{12}} \ \sigma^2(z_1,z_2)\,
\ee
and the normalised number-weighted halo bias within the bin is
.\ba
b(i_M,i_z) &= \frac{1}{N_\mr{cl}(i_M,i_z)} \int_{\substack{z\in \mr{bin}(i_z) \\ M\in \mr{bin}(i_M)}} \dd M \, \dd V \ \frac{\dd n_h}{\dd M} \ b(M,z)
.\ea
This approximation implicitly assumes that $b(M,z)$ varies slowly with redshift within the bin $i_z$, compared to $\sigma^2(z_1,z_2)$. \\
For a cylindrical window function of height $\delta r$ in the flat sky case, the sample variance matrix $S_{i_z,j_z}$ takes the form \citep{Hu2003,Lima2007,Aguena2016}
\ba
S&_{i_z,j_z} = \frac{1}{2\pi^2} \int k_\perp \, \dd k_\perp \; 4 \; \frac{J_1(k_\perp \theta_S r_1)}{k_\perp \theta_S r_1} \; \frac{J_1(k_\perp \theta_S r_2)}{k_\perp \theta_S r_2}\nonumber \\ \times &\int \dd k_\parallel \; j_0\left(\frac{k_\parallel  \delta r_1}{2}\right)  j_0\left(\frac{k_\parallel  \delta r_2}{2}\right)  \cos\left[k_\parallel (r_1-r_2)\right]  P_m(k|z_{12})\,,
\ea
where the power spectrum $P_m(k|z_{12})$ is evaluated at the centre of the respective redshift bins.
A nice feature of this approximation is that it removes the need to compute a double redshift integral. Furthermore the double integral over $(k_\perp,k_\parallel)$ only needs to be computed $n_z^2$ times (where $n_z$ is the number of redshift bins), instead of all redshift pairs $(z_1,z_2)$ needed to compute the redshift integral otherwise. This speeds up considerably the computation of SSC. We call this approximation the Sij method.

Second, another approximation used \citep[e.g.][]{Krause2016} is that $\sigma^2(z_1,z_2)$ is a Dirac delta function at $z_1=z_2$, so that the double redshift integral collapses into a single integral~
\ba
\nonumber \Cov^\mr{SSC}(&N_\mr{cl}(i_M,i_z),N_\mr{cl}(j_M,j_z)) \approx \nonumber \\ &\delta_{i_z,j_z} \int \dd V \, r^2(z) \, \frac{\partial n_h}{\partial \delta_b}(i_M,z) \, \frac{\partial n_h}{\partial \delta_b}(j_M,z) \, \sigma_b(\Omega_S,z)\,,
\ea
where, following notation from \citet{Krause2016}, we have
\be
\sigma_b(\Omega_S,z) = \int \frac{k_\perp \, \dd k_\perp}{2\pi} \ P_m(k_\perp,z) \ \left[\frac{2 J_1(k_\perp r(z) \theta_S)}{k_\perp r(z) \theta_S}\right]^2 \,,
\ee
where $P_m(k_\perp,z)=P_m(k=k_\perp,z)$, and we note that the $k_\parallel$ integral has disappeared and that $\sigma_b$ has units of Mpc/$h$.

The assumption behind this approximation is that the 3D window function $W(\xx)$ is much wider in the radial direction than in the transverse direction. Thus super-survey modes have $k_\parallel \ll k_\perp$, and $P_m(k)$ can be taken as approximately constant within the $k_\parallel$ integral of Eq.~(\ref{Eq:sigma2_flatsky}). We thus expect the approximation to fare better for wide redshift bins and small angles. By limiting the computation to equal redshift bins $i_z=j_z$, and reducing the multiple integrals to a single redshift and a single wave-vector $k_\perp$ integral, the approximation speeds up the SSC computation considerably. Hereafter, we call this approximation the KE method.

\begin{figure}[!th]
\begin{center}
\includegraphics[width=0.99\linewidth]{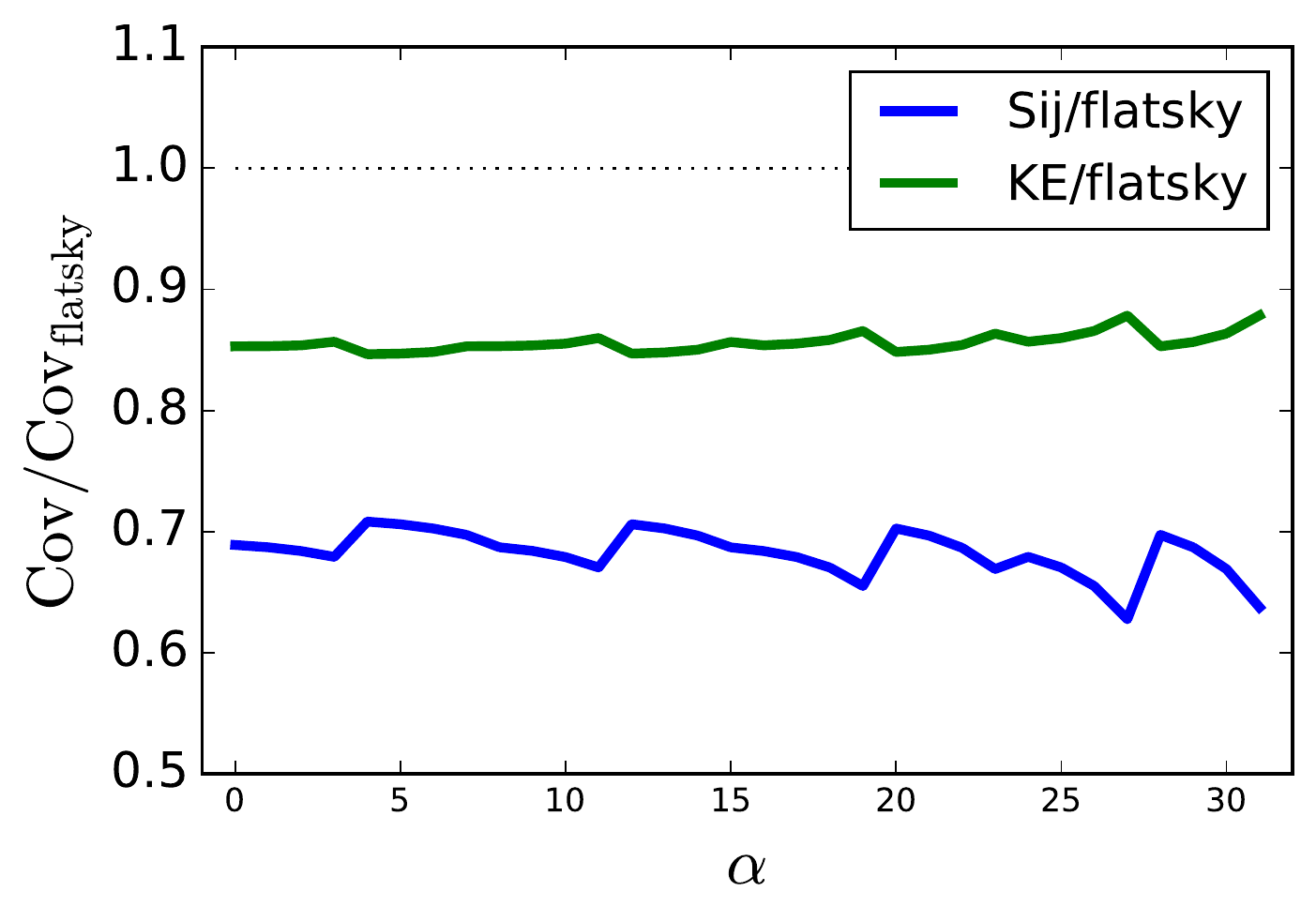}
\caption{Ratio of the SSC covariances from the Sij approximation and KE approximation to the full computation. Auto-z covariances are ordered as a function of $\alpha=j_M+n_M \, i_M+n_M^2 \, i_z$ (see text for details).}
\label{Fig:ratio_Sij-KE_overflatsky}
\end{center}
\end{figure}

We implemented all three SSC methods numerically: 1) full computation from Eqs.~(\ref{Eq:SSC_Ncl}) and (\ref{Eq:sigma2_flatsky}), 2) Sij approximation, and 3) KE approximation.
Figure \ref{Fig:ratio_Sij-KE_overflatsky} shows the ratio of the auto-z covariances to the full computation, plotted as a function of $\alpha=j_M+n_M \, i_M+n_M^2 \, i_z$; i.e. the first four points show the covariance ratio for $i_M=i_z=0$ ($14<\log M<14.5$ and $0.4<z<0.5$) and varying $j_M$, the next four points are for $i_M=1$, $i_z=0$, etc. This order is simply convenient to plotting this multi-variable function.
As visible in Fig. \ref{Fig:ratio_Sij-KE_overflatsky} both the Sij and KE approximations underpredict the amplitude of the covariance by $30-35\%$ and by $\sim15\%$, respectively.
The fact that the Sij approximation underpredicts the covariance conforms to our intuition; indeed, by taking values at the centre of the redshift bins, the approximation neglects the fact that the mass function and matter power spectrum both grow with time, which makes their integral larger\footnote{e.g. $1/3=\int_0^1 x^2 \dd x > 1/2*\int_0^1 x \dd x = 1/4$}. We checked that the Sij method and full computation are in much better agreement for smaller redshift bins, and indeed it can be seen analytically that they reduce to each other in the limit $\Delta z\rightarrow 0$.

\begin{figure*}[!th]
\begin{center}
\includegraphics[width=0.3\linewidth]{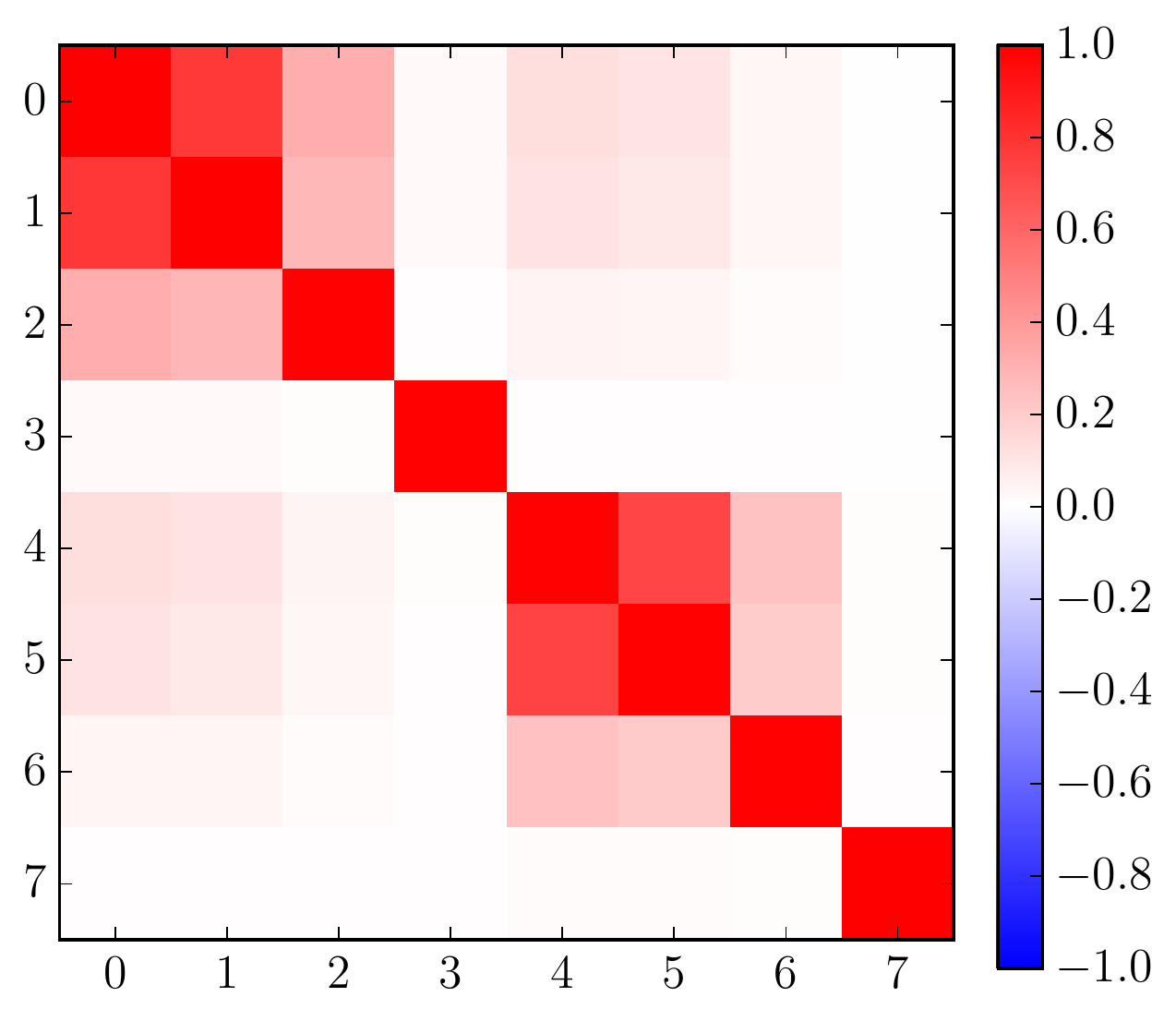}
\includegraphics[width=0.3\linewidth]{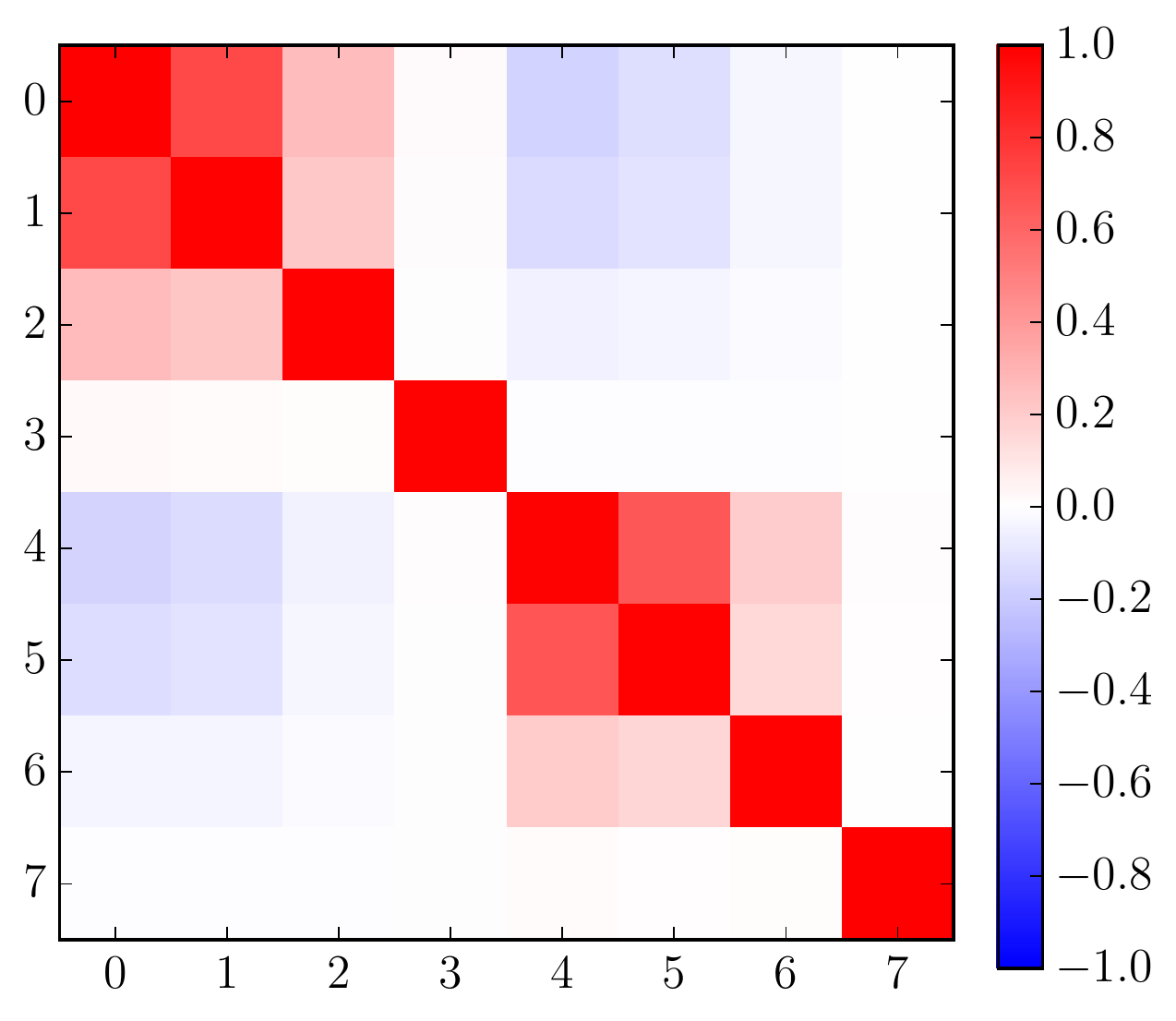} 
\includegraphics[width=0.3\linewidth]{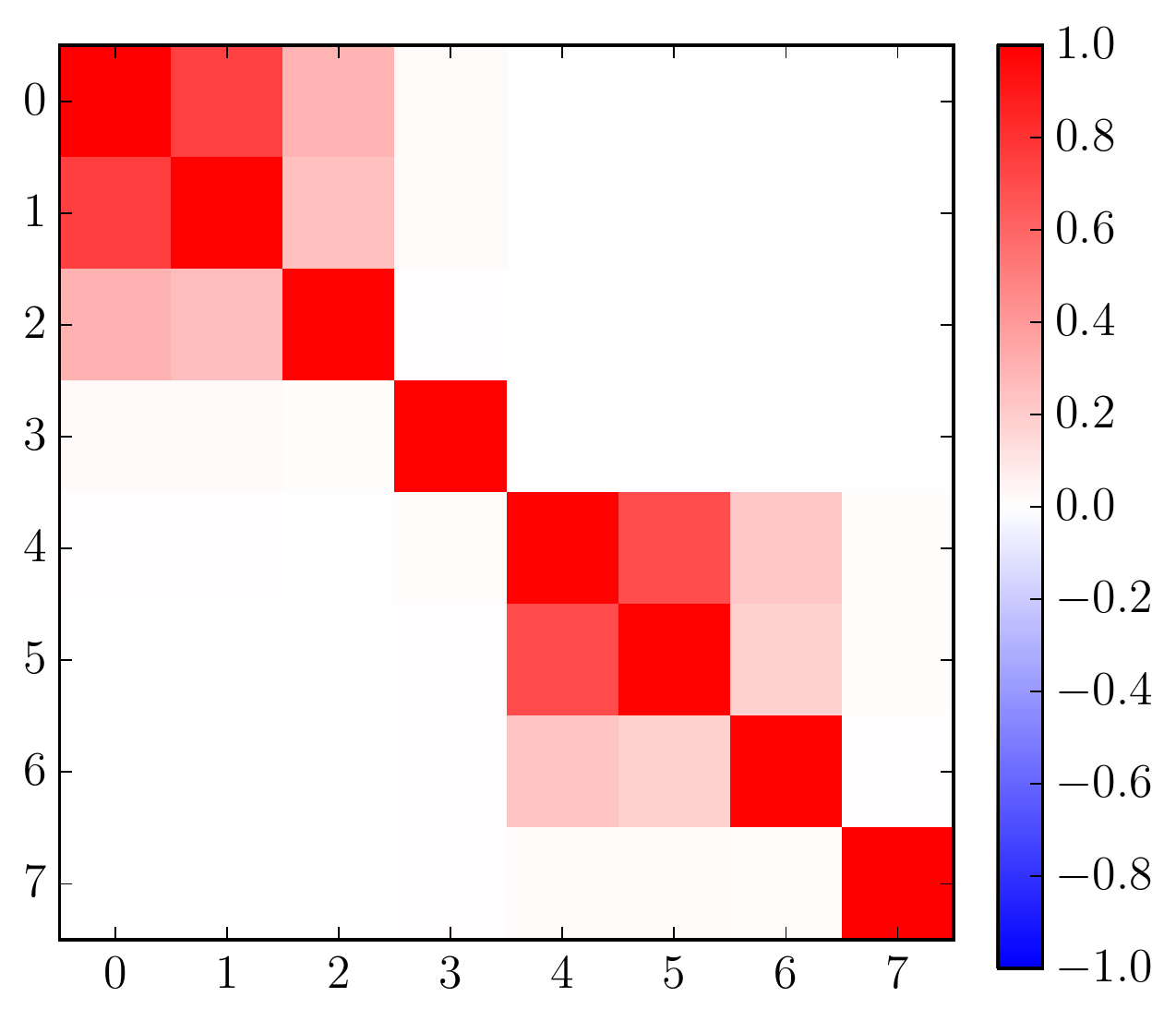}
\caption{Comparison of the cluster counts correlation matrix for different SSC computations, for a survey with angular radius $\theta_S=5$ deg. The SSC matrix is shown in 2 blocks for the redshift bins, and each block has 4 entries for the logarithmic mass bins.
\textit{From left to right :} full numerical computation from Eq.~(\ref{Eq:SSC_Ncl}), Sij approximation, and KE approximation.
}
\label{Fig:comp-SSC-approxs}
\end{center}
\end{figure*}

In Fig.~\ref{Fig:comp-SSC-approxs}, we show the correlation matrices for cluster counts obtained from each of the three SSC computations. The data points are ordered with increasing mass then increasing redshift, i.e. the number count index is $i_M+ n_M \, i_z$.\footnote{For example 0 refers to the lowest mass bin ($14<\log M<14.5$) in the first redshift bin ($0.4<z<0.5$), 3 refers to the highest mass bin ($15.5<\log M<16$) still in the first redshift bin, and then we go to the second redshift bin ($0.5<z<0.6$).}

As visible in Fig.~\ref{Fig:comp-SSC-approxs}, for cross-redshift covariances the full computation gives a positive correlation between our two redshift bins, reaching $\sim13\%$ at maximum for \mbox{$\Delta z=0.1$}. On the other hand, the Sij approximation predicts anti-correlation between the bins, reaching at maximum $-16\%$, while the KE approximation by construction predicts zero correlation. We checked that the Sij method performs better as the redshift bin width decreases. For instance at $\Delta z=0.01$, the agreement with the full computation is very good.

As a conclusion of this section, we see that the full computation is necessary to faithfully predict the SSC covariance matrix in the case $\Delta z=0.1$. In fact, this is a representative binning for current photometric galaxy surveys whose present photo-z errors are of this order \cite[e.g.][]{Sanchez2014}. 

%%%%%%%%%%%%%%%%%%%%%%%%%%%%%%%%%%%%%%%%%%%%%%%%%%%%%%%%%%%%%%%%%%%%
\section{Super sample covariance for partial sky coverage}\label{Sect:psky-derivation}

In this section, we assume for simplicity that the survey mask is independent of redshift, as we in fact already did implicitly in Eq.~(\ref{Eq:def-deltab}). It is however straightforward to generalise the formalism to a redshift-dependent mask (see Appendix \ref{App:maskz}).

%%%%%%%%%%%%%%%%%%%%
\subsection{Formal derivation}

We want to compute the covariance of the background mode
\be
\sigma^2(z_1,z_2) = \lbra \delta_b(z_1) \, \delta_b(z_2)\rbra \,.
\ee

Given that the mask has zero value outside the survey, and given the normalisation of spherical harmonics such that $Y_{00} = 1/\sqrt{4\pi}$, the background mode of the survey is related to the monopole of the masked matter density as~
\be
\delta_b(z) = \frac{1}{\sqrt{4\pi} \, f_\mr{SKY}} \ a_{00}^\mr{masked}(z)\,,
\ee
such that we have
\be
\sigma^2(z_1,z_2) = \frac{1}{4\pi \, f_\mr{SKY}^2} \ \hat{C}_0^\mr{masked}(z_1,z_2)\,,
\ee
where $\hat{C}_0^\mr{masked}$ is the angular power spectrum at $\ell=0$ of the masked matter density field. From $C_\ell$ pseudo-spectrum methods \citep[e.g.][]{Hivon2002}, we know that the masked power spectrum is related to the true spectrum via the so-called coupling matrix, such that~
\be
\hat{C}_0^\mr{masked}(z_1,z_2) = \sum_\ell \mathcal{M}_{0,\ell} \, C_\ell^\mr{true}(z_1,z_2)\,,
\ee
where $C_\ell^\mr{true}$ (hereafter simply $C_\ell^\mr{m}$) is the angular spectrum of matter density in an infinitesimal redshift shell. The coupling matrix is given by~
\be
\mathcal{M}_{\ell_1,\ell_2} = (2\ell_2+1) \sum_{\ell_3} \frac{(2\ell_3+1)}{4\pi} \ \threeJz{\ell_1}{\ell_2}{\ell_3}^2 \ C_{\ell_3}(W)\,,
\ee
where $C_{\ell}(W)$ is the angular power spectrum of the angular window $W(\hn)$.
In our case, given the properties of Wigner symbols, this simplifies to~\be
\mathcal{M}_{0,\ell} = \frac{(2\ell+1)^2}{4\pi} \ \threeJz{0}{\ell}{\ell}^2 \ C_{\ell}(W)\,.
\ee

Furthermore we have $\threeJz{0}{\ell}{\ell}=({2\ell+1})^{-1/2}$ and thus
\ba\label{Eq:sigma2-psky}
\sigma^2(z_1,z_2) &= \frac{1}{\Omega_\mr{S}^2} \sum_\ell (2\ell+1) \ C_\ell(W) \ C_\ell^\mr{m}(z_1,z_2)\,,
\ea
where $\Omega_\mr{S}=4\pi \, f_\mr{SKY}$ and the matter angular power spectrum $C_\ell^\mr{m}$ is given by \citep[e.g.][]{Campagne2017}
\be\label{Eq:Cl-matter}
C_\ell^\mr{m}(z_1,z_2) = \frac{2}{\pi} \int k^2 \, \dd k \; j_\ell(k r_1) \, j_\ell(k r_2) \; P_m(k|z_{12})\,.
\ee
The combined Eqs.~(\ref{Eq:sigma2-psky}) and (\ref{Eq:Cl-matter}) represent the main analytical result of this article.

%%%%%%%%%%%%%%%%%%%%
\subsection{Limiting cases and remarks}

It is interesting to examine a few limiting cases 
of Eqs.~(\ref{Eq:sigma2-psky}) and (\ref{Eq:Cl-matter}). 
In the full sky case, we have $C_\ell(W) = 4\pi \, \delta_{0,\ell}$ and $f_\mr{SKY}=1$, and therefore thus we have
\be
\sigma^2(z_1,z_2) = \sigma^2_\mr{fullsky}(z_1,z_2) = \left. C_0^\mr{m}(z_1,z_2) \ \middle/ \ 4\pi \right. \,,
\ee
recovering indeed Eq.~(\ref{Eq:sigma2_fullsky}).

In partial sky, $C_0(W) = 4\pi \, f_\mr{SKY}^2$, such that
\be
\sigma^2(z_1,z_2) = \sigma^2_\mr{fullsky}(z_1,z_2) + \frac{1}{\Omega_\mr{S}^2} \sum_{\ell\geq 1} (2\ell+1) \ C_\ell(W) \ C_\ell^\mr{m}(z_1,z_2)\,,
\ee
i.e. the full sky covariance term is the first in the sum contributing to the partial-sky covariance.

If $C_\ell^\mr{m}(z_1,z_2)$ is scale independent, i.e. 
$C_\ell^\mr{m}(z_1,z_2)=C_0^\mr{m}(z_1,z_2)$ independently of $\ell$ (e.g. for $P_m(k|z_{12})$=constant, see Sect.~\ref{Sect:recover-flatsky}), we have 
\ba
\sigma^2(z_1,z_2)&=\sigma^2_\mr{fullsky}(z_1,z_2) \times \frac{1}{f_\mr{SKY}^2}\sum_\ell \frac{2\ell+1}{4\pi} \, C_\ell(W) \nonumber \\
&=\sigma^2_\mr{fullsky}(z_1,z_2)/f_\mr{SKY}\,,
\ea
and therefore
$\Cov^\mr{SSC}=\Cov^\mr{SSC}_\mr{fullsky}/f_\mr{SKY}$, i.e. we obtain the usual $f_\mr{SKY}$ approximation to partial sky covariance. Conversely the reverse is also true: if the $f_\mr{SKY}$ approximation holds for any mask, then $C_\ell^\mr{m}(z_1,z_2)$ is constant.
Given that for viable cosmological models $C_\ell^\mr{m}(z_1,z_2)$ is not constant, this shows that SSC is a non-trivial source of covariance that cannot be treated by classical approximations.

Now we notice that Eq.~(\ref{Eq:sigma2-psky}) can be rewritten as
\be \label{Eq:sigma2-psky-rewritten}
\sigma^2(z_1,z_2) = \frac{1}{f_\mr{SKY}^2} \sum_\ell \frac{2\ell+1}{4\pi} \ C_\ell(W) \ \sigma_\ell^2(z_1,z_2)\,,
\ee
with
\ba \label{Eq:sigma_ell}
\sigma_\ell^2(z_1,z_2) &= \frac{C_\ell^\mr{m}(z_1,z_2)}{4\pi} \nonumber \\
&= \frac{1}{2\pi^2} \int k^2 \, \dd k \; j_\ell(k r_1) \, j_\ell(k r_2) \; P_m(k|z_{12})\,,
\ea
such that in full sky $\sigma^2(z_1,z_2)=\sigma_0^2(z_1,z_2)$.
Applying this to the SSC covariance of two observables, for example number counts, we have
\be\label{Eq:covSSC-sum-covell}
\Cov^\mr{SSC}(\mathcal{O}_1,\mathcal{O}_2) = \frac{1}{f_\mr{SKY}^2} \sum_\ell \frac{2\ell+1}{4\pi} \ C_\ell(W) \ \Cov^\mr{SSC}_\ell(\mathcal{O}_1,\mathcal{O}_2)\,,
\ee
where
\ba\label{Eq:def-Cov_ell^SSC}
\Cov^\mr{SSC}_\ell(\mathcal{O}_1,\mathcal{O}_2) &= \int \dd V_{12} \ \frac{\partial \mathfrak{o}_1}{\partial \delta_b} \ \frac{\partial \mathfrak{o}_1}{\partial \delta_b} \ \sigma_\ell^2(z_1,z_2) \\
&= \frac{1}{4\pi}\int \dd V_{12} \ \frac{\partial \mathfrak{o}_1}{\partial \delta_b} \ \frac{\partial \mathfrak{o}_1}{\partial \delta_b} \ C_\ell^\mr{m}(z_1,z_2)\,. \nonumber
\ea

Therefore the combined Eqs.~(\ref{Eq:sigma2-psky-rewritten}), (\ref{Eq:sigma_ell}), (\ref{Eq:covSSC-sum-covell}), (\ref{Eq:def-Cov_ell^SSC}), along with the window spectrum $C_\ell(W)$, allow for the computation of the  SSC in partial sky for a general window and binning choice.

Numerically, we can pre-compute and tabulate $\Cov^\mr{SSC}_\ell(\mathcal{O}_1,\mathcal{O}_2)$, and then just change $C_\ell(W)$ as the mask varies. This enables easier studies of the mask effect such as optimisations of survey strategy \citep{Takahashi2014}, forecasts for improvements as a survey area grows, and comparisons between different surveys. Conversely, for a fixed survey with a well-defined mask, we can pre-compute $C_\ell(W)$ for the specific survey mask, and change $\Cov^\mr{SSC}_\ell(\mathcal{O}_1,\mathcal{O}_2)$ as a function of cosmology within likelihood cosmological analyses, taking full account  of geometry, mask, selection, and cosmological dependencies in the covariances, and therefore deriving reliable parameter uncertainties. 

\begin{figure*}[!th]
\begin{center}
\includegraphics[width=0.32\linewidth]{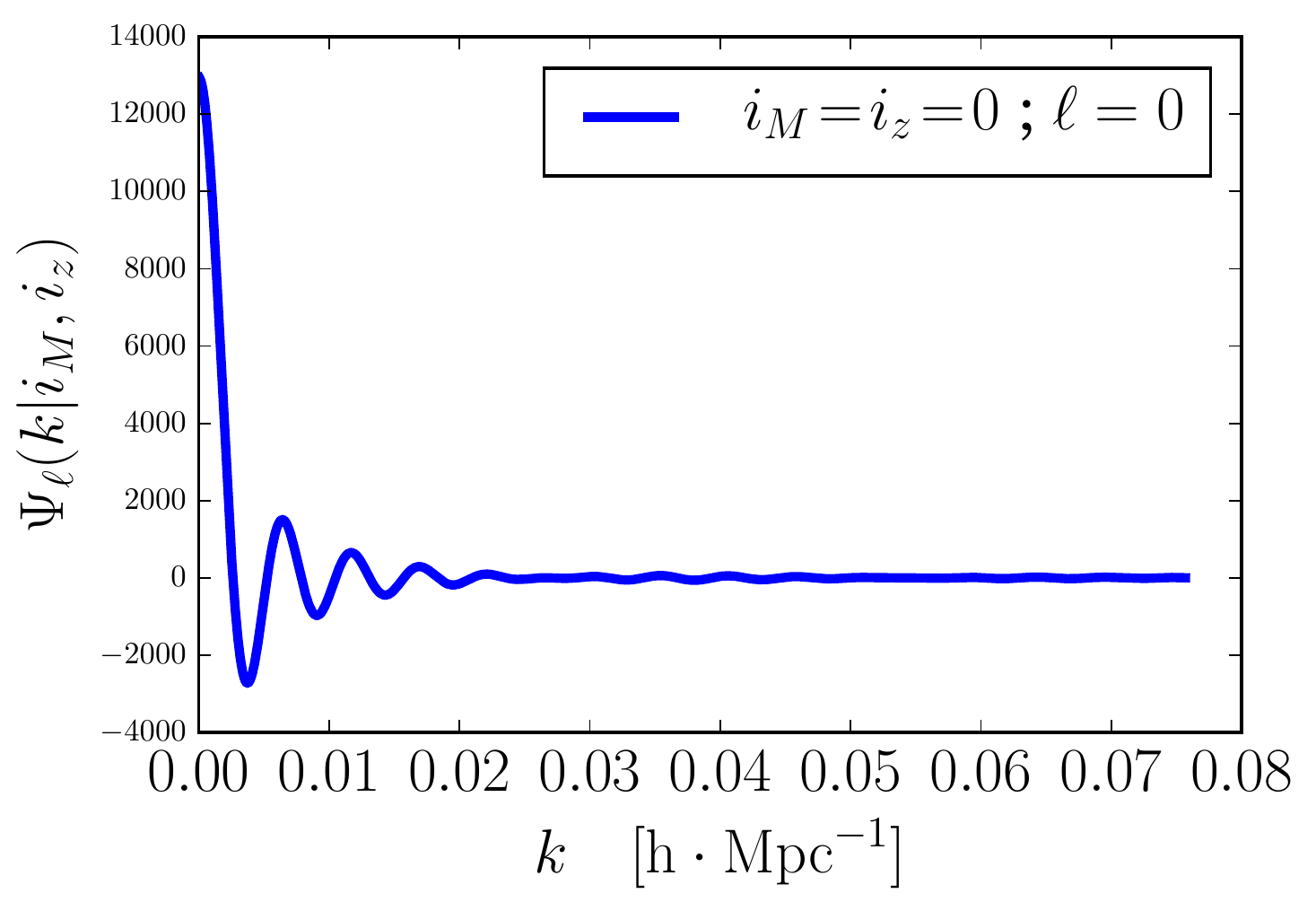}
\includegraphics[width=0.32\linewidth]{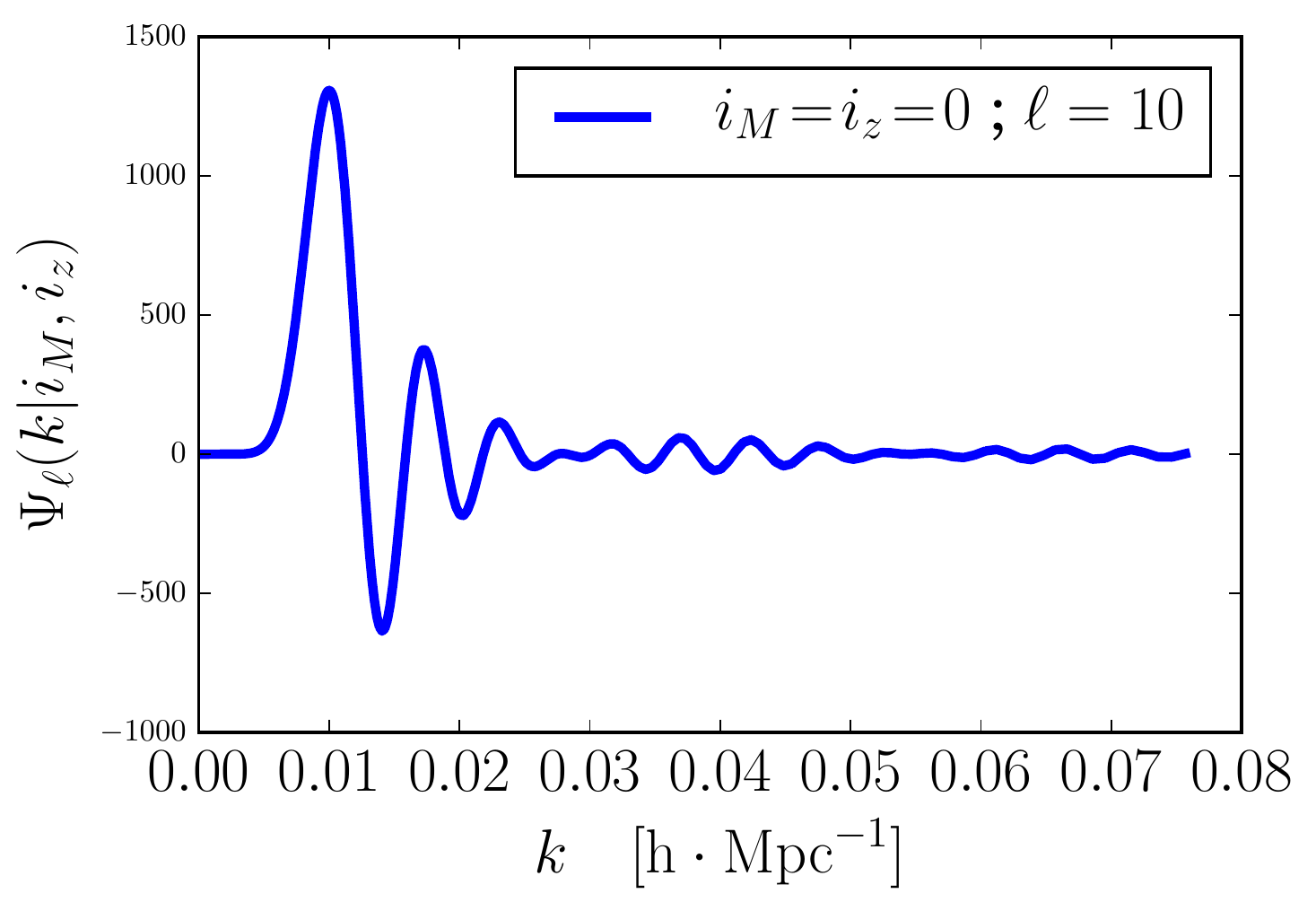}
\includegraphics[width=0.32\linewidth]{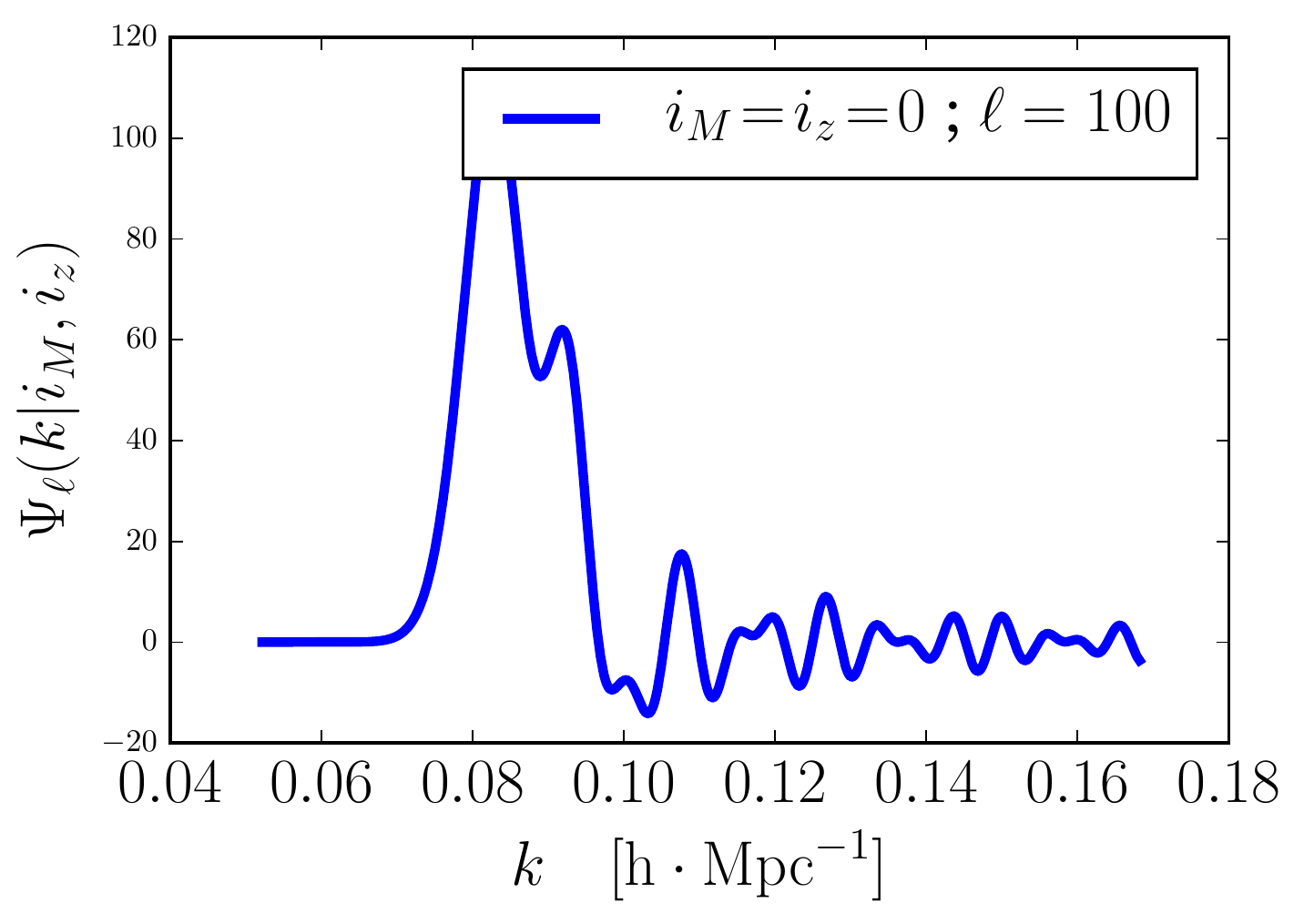}
\caption{$\Psi^{n_h}_\ell(k|i_M,i_z)$ as a function of k, for the representative case $i_M=i_z=0$ ($\log M=14-14.5$, $z=0.4-0.5$). \textit{Left:} at the lowest multipole $\ell=0$. \textit{Centre:} at $\ell=10$. \textit{Right:} at $\ell=100$.}
\label{Fig:Psi_l_nh}
\end{center}
\end{figure*}

%%%%%%%%%%%%%%%%%%%%
\subsection{Implementation}\label{Sect:implementation}

When estimating the SSC from Eq.~(\ref{Eq:covSSC-sum-covell}), one numerical difficulty is the evaluation of the integral in Eq.~(\ref{Eq:sigma_ell}), given that the Bessel functions highly oscillate with slow damping as $kr_i\rightarrow \infty$.

For the first multipoles, we may express the Bessel functions in terms of sine and cosine. Through trigonometrical identities, the integrals with products of Bessel function can thus be expressed as a sum of Fourier transforms. Derivations and expressions for the first three multipoles are given in Appendix \ref{App:first-multipoles}. For illustrative purposes we give below the expression for $\ell=0$
\be
C_0^\mr{m}(z_1,z_2) = \frac{I_0^c(r_1-r_2) - I_0^c(r_1+r_2)}{2}\,,
\ee
where
\be
I_0^c(r) = \frac{2}{\pi} \int k^2 \, \dd k \; \cos(kr) \; P_m(k|z_{12})/k^2
\ee
is a continuous cosine transform that can be efficiently approximated numerically with a discrete fast Fourier transform (FFT).
However, as argued in Appendix \ref{App:first-multipoles}, this method becomes too cumbersome at high $\ell$ and may also become numerically unstable. We could carry it out only for $\ell=0,1,2$.

As an alternative, we decided not to evaluate $\sigma^2(z_1,z_2)$, but instead to switch the order of the integrals over $k$ and $z$,  to compute the covariance directly as~
\ba\label{Eq:covellSSC-intk}
\nonumber\Cov^\mr{SSC}_\ell(N_\mr{cl}(i_M,i_z),N_\mr{cl}(j_M,j_z)) &= \frac{1}{2\pi^2} \int k^2 \dd k \; P_m(k|z=0) \\
& \hspace{-0.7cm}\times \Psi^{n_h}_\ell(k|i_M,i_z) \ \Psi^{n_h}_\ell(k|j_M,j_z)\,,
\ea
where the kernel $\Psi^{n_h}_\ell$ is given by 
\be \label{Eq:Psi_ell}
\Psi^{n_h}_\ell(k|i_M,i_z) \equiv \int_{z\in\mr{bin}(i_z)} \dd V \, G(z) \, \frac{\partial n_h}{\partial \delta_b}(i_M,z) \, j_\ell(k r)\,.
\ee

As can be seen in Fig.~\ref{Fig:Psi_l_nh}, this redshift integral effectively damps out the Bessel oscillations on scales $k>k_\mr{peak}+2\pi/\Delta r(i_z),$ where $\Delta r(i_z)=r(z_\mr{max})-r(z_\mr{min})$ is the width of the redshift bin in terms of comoving distance. This damping makes the $k$ integral Eq.~(\ref{Eq:covellSSC-intk}) much easier to deal with numerically compared to Eq.~(\ref{Eq:Cl-matter}), since the integrand support is now more compact. So numerically we could just compute the later integrals by brute force, and this is the method we use for all numerical results shown hereafter in this article. We note that this method does not compute $\sigma^2(z_1,z_2)$ as an intermediary product, and thus prevents from comparison of this quantity in partial sky with its alter ego in the full sky or flat sky limits, as in Fig.~\ref{Fig:sigma2-flatvsfull}.

We remark that alternatively, Eq.~(\ref{Eq:Psi_ell}) can be expressed as a Hankel transform
\be \label{Eq:Psi_ell_Hankel}
\Psi^{n_h}_\ell(k|i_M,i_r) = \int_{r\in\mr{bin}(i_r)} \dd r \ r^2 \, G(r) \, \frac{\partial n_h}{\partial \delta_b}(i_M,r) \, j_\ell(k r)\,,
\ee
via the inversion of the $r(z)$ relation and defining a radial bin $i_r$ from $i_z$. Eq.~(\ref{Eq:Psi_ell_Hankel}) can be efficiently evaluated using Fourier transform methods such as FFTLog \citep[e.g.][]{Hamilton2000, Camacho_etal_inprep}. Although this is not the numerical method we used in this article, we note it can be a useful approach for future high-precision applications.

The results presented in this article consider the case of cluster counts,  and the equations can be straightforwardly generalised to other probes of the LSS; see Appendix \ref{App:other-probes}.

%%%%%%%%%%%%%%%%%%%%%%%%%%%%%%%%%%%%%%%%%%%%%%%%%%%%%%%%%%%%%%%%%%%%
\section{Results}\label{Sect:results}

First, as a consistency test, we checked $\Cov^\mr{SSC}_{\ell=0}$ against the full sky covariance matrix computed via $\sigma^2_\mr{fullsky}(z_1,z_2)$ given in Eq.~(\ref{Eq:sigma2_fullsky}) \citep[e.g.][]{Lacasa2016}. We find good agreement, to 0.8\% precision on the auto-redshift covariance, and 7\% precision on the cross-redshift covariance.

\begin{figure}[!th]
\begin{center}
\includegraphics[width=\linewidth]{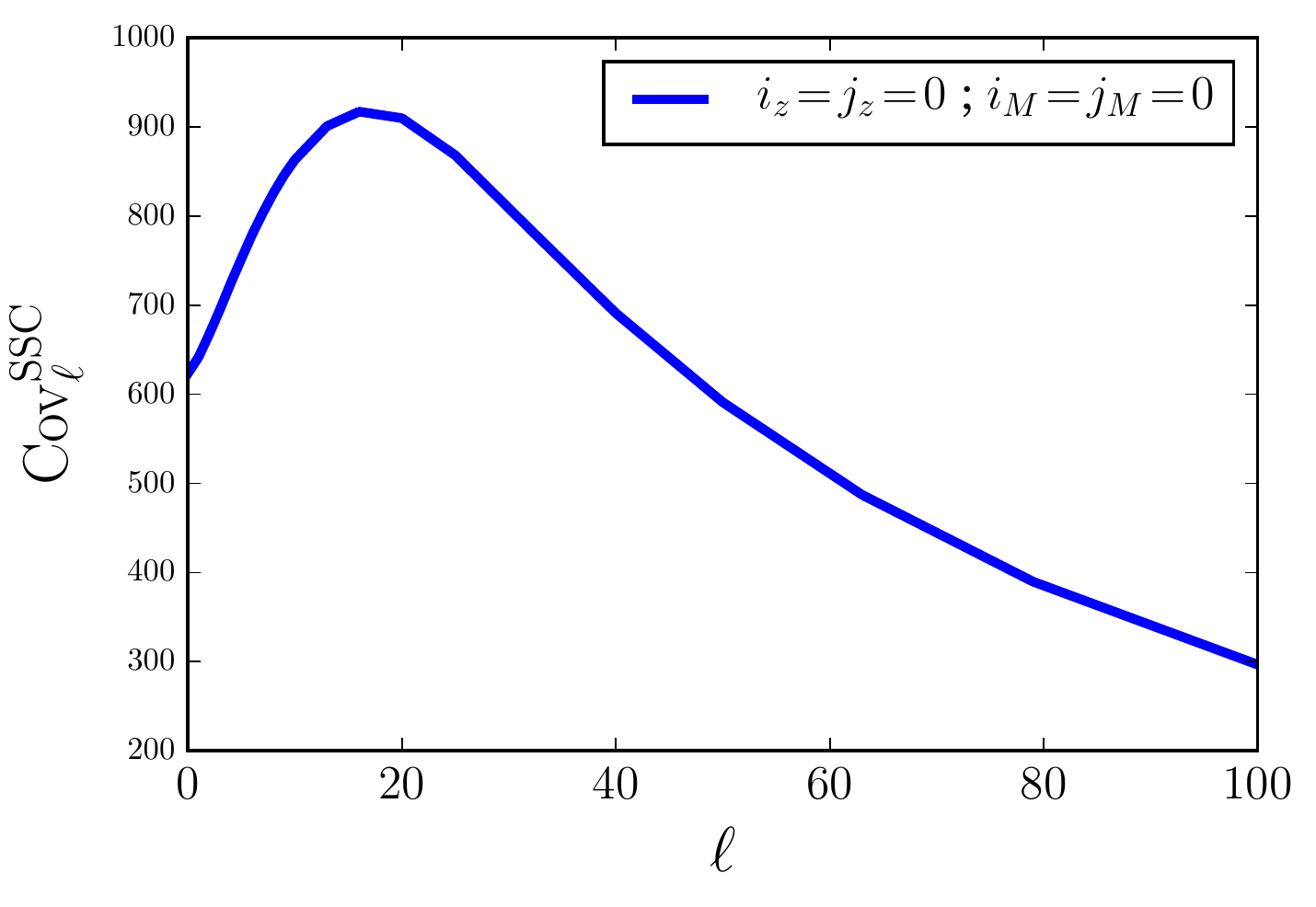}
\includegraphics[width=\linewidth]{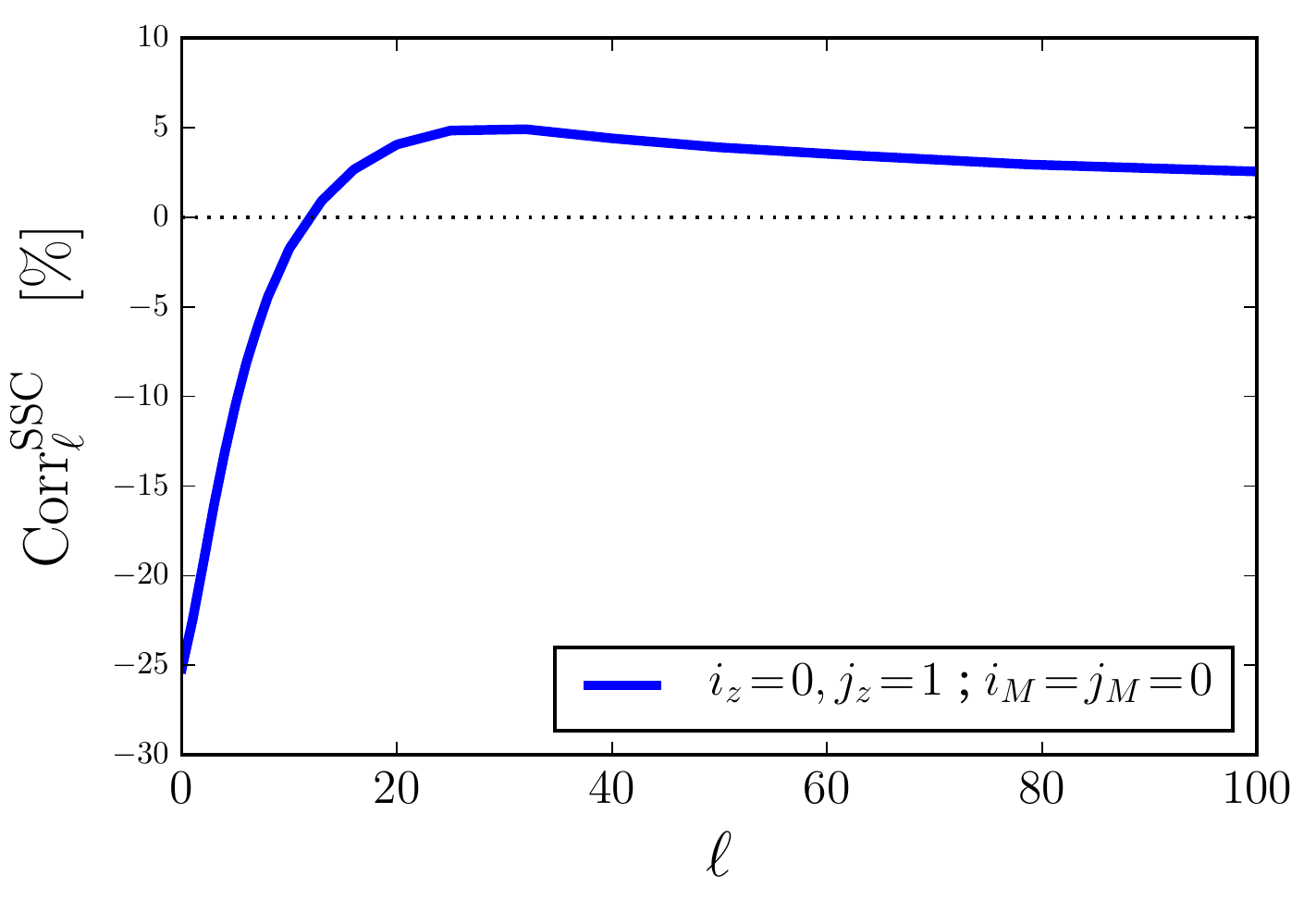}
\caption{$\Cov^\mr{SSC}_\ell$ as a function of $\ell$, in the representative case $i_M=j_M=0$ ($\log M=14-14.5$). \textit{Top:} same redshift $i_z=j_z=0$ ($z=0.4-0.5$). \textit{Bottom:} cross-redshift $i_z=1 \times j_z=2$ ($z=[0.4-0.5]\times[0.5-0.6]$).}
\label{Fig:cov_ell_SSC_vs_ell}
\end{center}
\end{figure}

Second, we show in Fig.~\ref{Fig:cov_ell_SSC_vs_ell} the general results for $\Cov^\mr{SSC}_\ell(N_\mr{cl}(i_M,i_z),N_\mr{cl}(j_M,j_z))$ in two representative cases: auto-covariance of the same redshift bin and cross-correlation (i.e. cross-covariance normalised by the corresponding auto-covariances) between two redshift bins. Both plots are for the lowest mass bin ($i_M=j_M=0$, corresponding to $\log M=14-14.5$), although the shape of the curves does not change significantly when taking other mass bins (even with $i_M\neq j_M$), only their amplitude changes.

For the auto-covariance (left plot), we see that it first rises with $\ell$, to a maximum corresponding to the angular scale of the matter-radiation equality (i.e. the peak of $P_m(k)$) at that redshift, and then decreases monotonically. This behaviour conforms to our expectations, since the scale dependence is that of a projection of the 3D power spectrum $P_m(k)$. This means that $C_\ell^m(z_1,z_2)$ is not constant with $\ell$, and thus, as already discussed in Sect. \ref{Sect:psky-derivation}, it means that the $f_\mr{SKY}$ approximation does not hold for SSC.

For the cross-covariance (right plot), the situation is interesting as we see that the covariance is first negative,  increases, crosses zero, and reaches a maximum towards $\ell=25$, and then decreases asymptotically to zero. This means that small surveys have a robustly negligible covariance between redshift bins (at least for this bin width $\Delta z=0.1$), and can thus use block-diagonal covariance matrices to speed up their likelihoods estimations. However for surveys with a large sky coverage, the cross-covariance is non-negligible and depends on the survey area and shape, becoming either positive or negative depending on the mask. In those cases, careful estimation of the SSC is thus critical.

%%%%%%%%%%%%%%%%%%%%
\subsection{Application to a realistic survey mask}\label{Sect:applic-DES}

\begin{figure}[!th]
\begin{center}
\includegraphics[width=\linewidth]{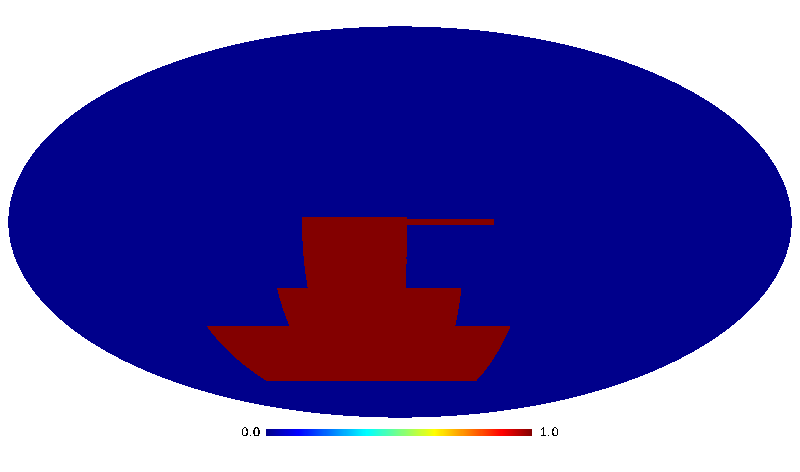}
\includegraphics[width=\linewidth]{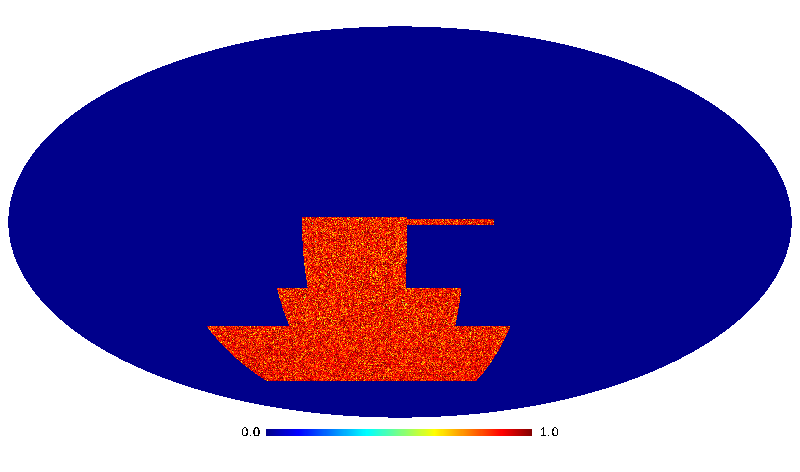}
\caption{Masks used in the analysis. \textit{Top:} simple footprint, assumed observed uniformly. \textit{Bottom:} same but with simulated 15\% rejection of observations due to systematics (see text for details).}
\label{Fig:masks}
\end{center}
\end{figure}

In order to illustrate the SSC method in a realistic case, we used two {\tt Healpix} \citep{Gorski2005} masks visible in Fig. \ref{Fig:masks}. The first mask is binary and is broadly similar to the footprint of DES\footnote{\url{www.darkenergysurvey.org}}\footnote{This mask was created and kindly provided to us by Flavia Sobreira.}, although we warn that it does not in any way represent the actual DES survey area, and we do not attempt to draw any particular conclusion for that survey.
The second mask represents a more pessimistic case, where we considered that 15\% of the survey area had to be discarded due to bright stars, satellite trails, or other systematics. In order to simulate this effect, we simply upgraded the mask to high resolution ({\tt nside}=4096), poked random holes in it, then degraded the mask back to the original resolution ({\tt nside}=1024).

\begin{figure}[!th]
\begin{center}
\includegraphics[width=\linewidth]{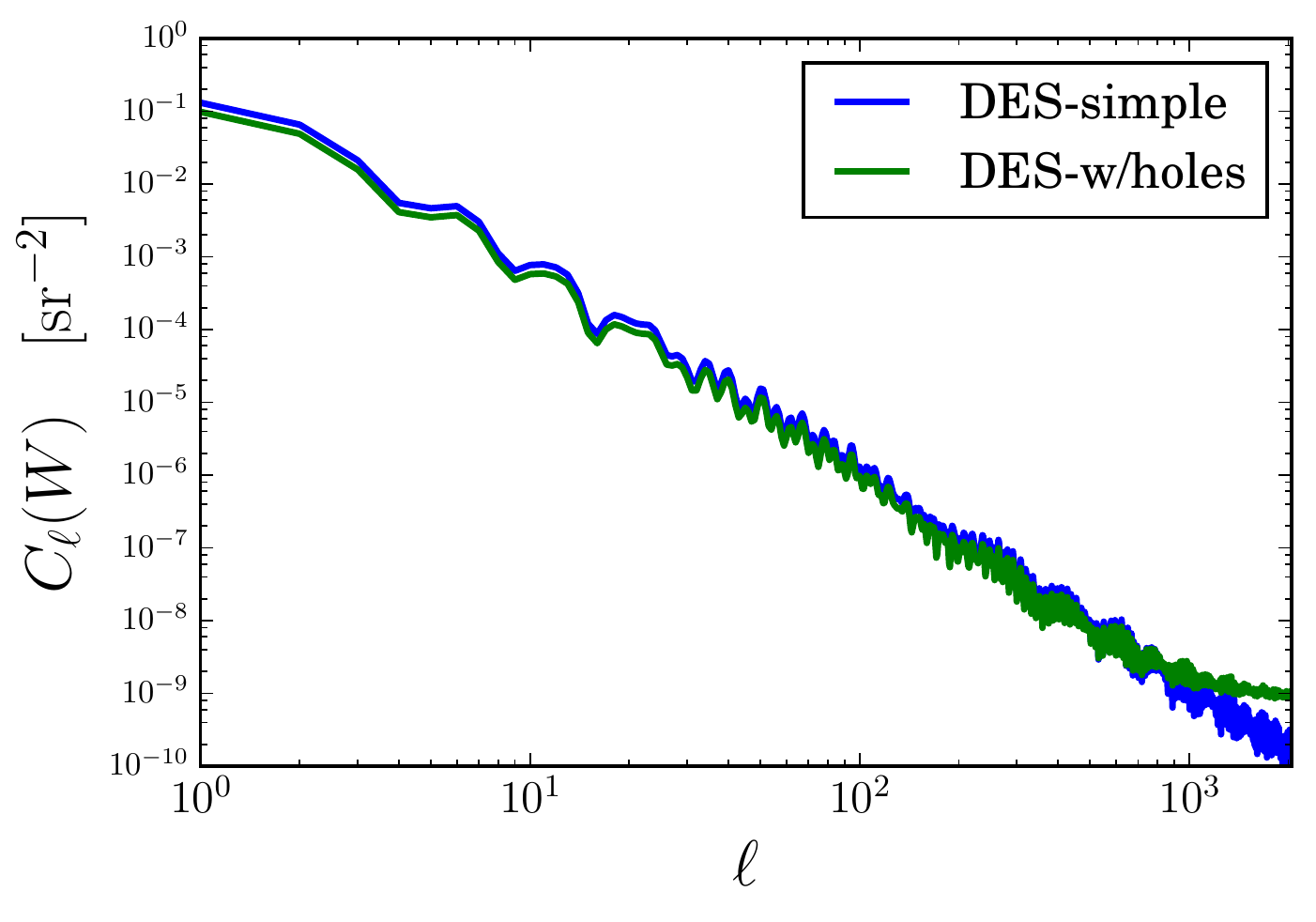}
\caption{Angular power spectra of the two masks used in the analysis and shown in Fig. \ref{Fig:masks}.}
\label{Fig:clmasks}
\end{center}
\end{figure}

The angular power spectra $C_\ell(W)$ for these two masks can be seen on Fig.~\ref{Fig:clmasks}.
At low multipoles, the two spectra differ by a constant multiplicative factor $0.85^2$, which is simply the ratio of the respective sky coverages. A second component in the spectrum of the second mask appears on smaller scales, which is a constant shot noise due to the random holes. However we see that this component is very subdominant, hence we can already expect that the only difference between the covariance of the two masks is due to the different sky coverage.

\begin{figure}[!th]
\begin{center}
\includegraphics[width=\linewidth]{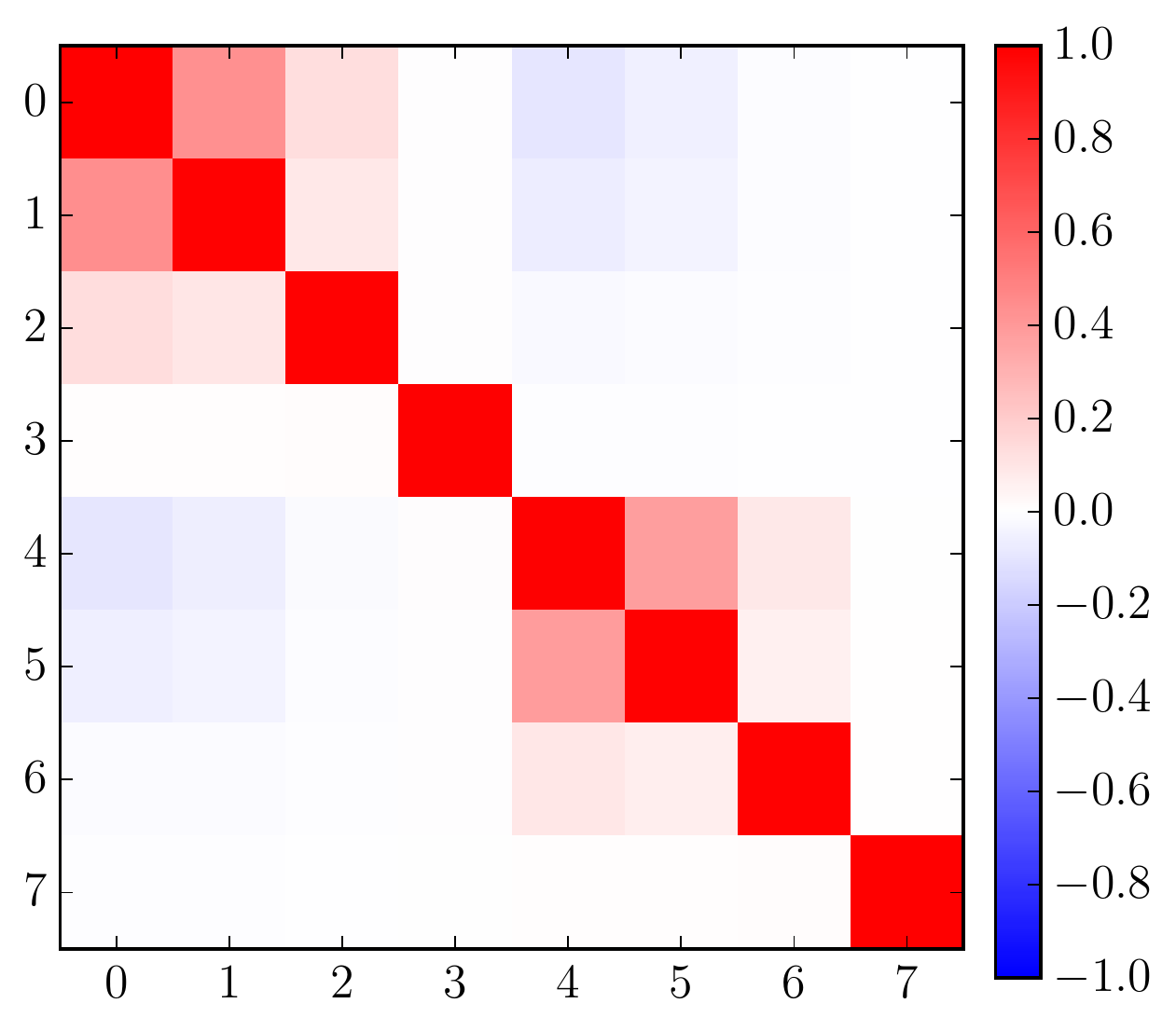}
\caption{Total correlation matrix (SSC + shot-noise) of cluster counts for the first mask. The matrix is organised, as in Fig. \ref{Fig:comp-SSC-approxs}, in two redshift blocks of increasing mass.}
\label{Fig:correl-mat-DESsimple}
\end{center}
\end{figure}

When implementing the sum over multipoles of Eq.~(\ref{Eq:covSSC-sum-covell}), we find that we reached 1\% convergence already at $\ell_\mr{max}=50$. This means that $\Cov_\ell^\mr{SSC}$ need only be computed  on a small number of multipoles for current and future surveys with large sky coverage, rendering the partial sky formalism developed here even more computationally efficient.
Comparing the SSC covariances of the two masks, we find that they indeed only differ due to the different sky fraction, but when renormalised by $f_\mr{SKY}$ they are identical to numerical precision (0.01\% in our case). The total correlation matrix (including shot-noise) for the first mask is shown in Fig.~\ref{Fig:correl-mat-DESsimple}. 

We see that the SSC is an important contribution to the covariance matrix, dominating the error bars at \mbox{$\log M \leq 14.5$}. Concerning cross-redshifts, the SSC yields an anti-correlation between the adjacent bins reaching up to -10\%. The cross-covariance is thus far from negligible for such a large survey.

The formalism presented in Sect. \ref{Sect:psky-derivation} is thus perfectly adapted to the numerical prediction of SSC, even in the case of a complex survey geometry. In fact we see that such computation is indeed necessary to reproduce the non-trivial behaviour of SSC, yielding, for example non-negligible anti-correlation of redshift bins, in the case presented above.

%%%%%%%%%%%%%%%%%%%%
\subsection{Flat sky limit}\label{Sect:recover-flatsky}

One remaining question is that of the link between the partial sky approach developed in Sect. \ref{Sect:psky-derivation} and the flat sky approximation used previously in the literature and shown in Eq.~(\ref{Eq:sigma2_flatsky}).
The two approaches cannot be compared or related at the level of the formalism because the $(k_\perp,k_\parallel)$ splitting does not apply for sufficiently large angles. However, we can compare what covariance the two formalisms predict in some limits.

First, we can compare the equations analytically in the case of a constant spectrum $P_m(k|z_{12})=\mr{cst}\equiv 1$. In the partial sky formalism we have
\ba
C_\ell^m(z_1,z_2) = \frac{2}{\pi} \int k^2 \dd k j_\ell(k r_1) j_\ell(k r_2) \times 1= \frac{\delta(r_1 - r_2)}{r_1^2} \,,
\ea
and
\ba
\Cov_\ell^\mr{SSC}(N_\mr{cl}(i_M,i_z),&N_\mr{cl}(j_M,j_z)) = \nonumber \\ & \frac{\delta_{i_z,j_z}}{4\pi} \!\int\! \dd V \, \frac{\partial n_h}{\partial \delta_b}(i_M,z) \, \frac{\partial n_h}{\partial \delta_b}(j_M,z) 
\ea
is independent of $\ell$. Thus the SSC is given by
\ba
\nonumber \Cov^\mr{SSC}\left(\cdots\right) &=
\frac{\delta_{i_z,j_z}}{4\pi} \!\int\! \dd V \, \frac{\partial n_h}{\partial \delta_b}(i_M,z) \, \frac{\partial n_h}{\partial \delta_b}(j_M,z)\\
\nonumber & \quad \times \frac{1}{f^2_\mr{SKY}} \underbrace{\sum_\ell \frac{(2\ell+1)}{4\pi} C_\ell(W)}_{=f_\mr{SKY}}\\
&= \frac{\delta_{i_z,j_z}}{\Omega_S} \int \dd V \ \frac{\partial n_h}{\partial \delta_b}(i_M,z) \, \frac{\partial n_h}{\partial \delta_b}(j_M,z)\,,
\label{Eq:CovSSC-const-spec}
\ea
where we made the counts $N_{\rm cl}$ implicit for the sake of clarity.

Now in the flat sky formalism we have\ba
\nonumber \sigma^2(z_1,z_2) &= \frac{1}{2\pi^2} \int_0^\infty k_\perp \, \dd k_\perp \, 4 \frac{J_1(k_\perp \theta_S r_1)}{k_\perp \theta_S r_1} \frac{J_1(k_\perp \theta_S r_2)}{k_\perp \theta_S r_2} \nonumber \\ 
&\qquad \times 
\underbrace{\int_0^\infty \dd k_\parallel \; \cos\left[k_\parallel (r_1-r_2)\right] \times 1}_{=(2\pi) \,\delta(r_1-r_2)/2} \nonumber \\
&=\frac{2}{\pi} \ \delta(r_1-r_2) \ \frac{1}{(\theta r_1)^2} \ \underbrace{\int x \, \dd x \ \left(J_1(x)/x\right)^2}_{=1/2} \nonumber \\
&= \frac{\delta(r_1 - r_2)}{r_1^2} \times \frac{1}{\Omega_S}\,,
\ea
where we changed variables $x=k_\perp \theta_S r_1$  and recall \mbox{$\Omega_S \approx \pi \theta_S^2$}.
Inserting this into Eq.~(\ref{Eq:SSC_Ncl}), we again obtain Eq.~(\ref{Eq:CovSSC-const-spec}). 
Therefore the two approaches indeed agree in the flat sky limit for a constant spectrum.

\begin{figure}[!th]
\begin{center}
\includegraphics[width=0.99\linewidth]{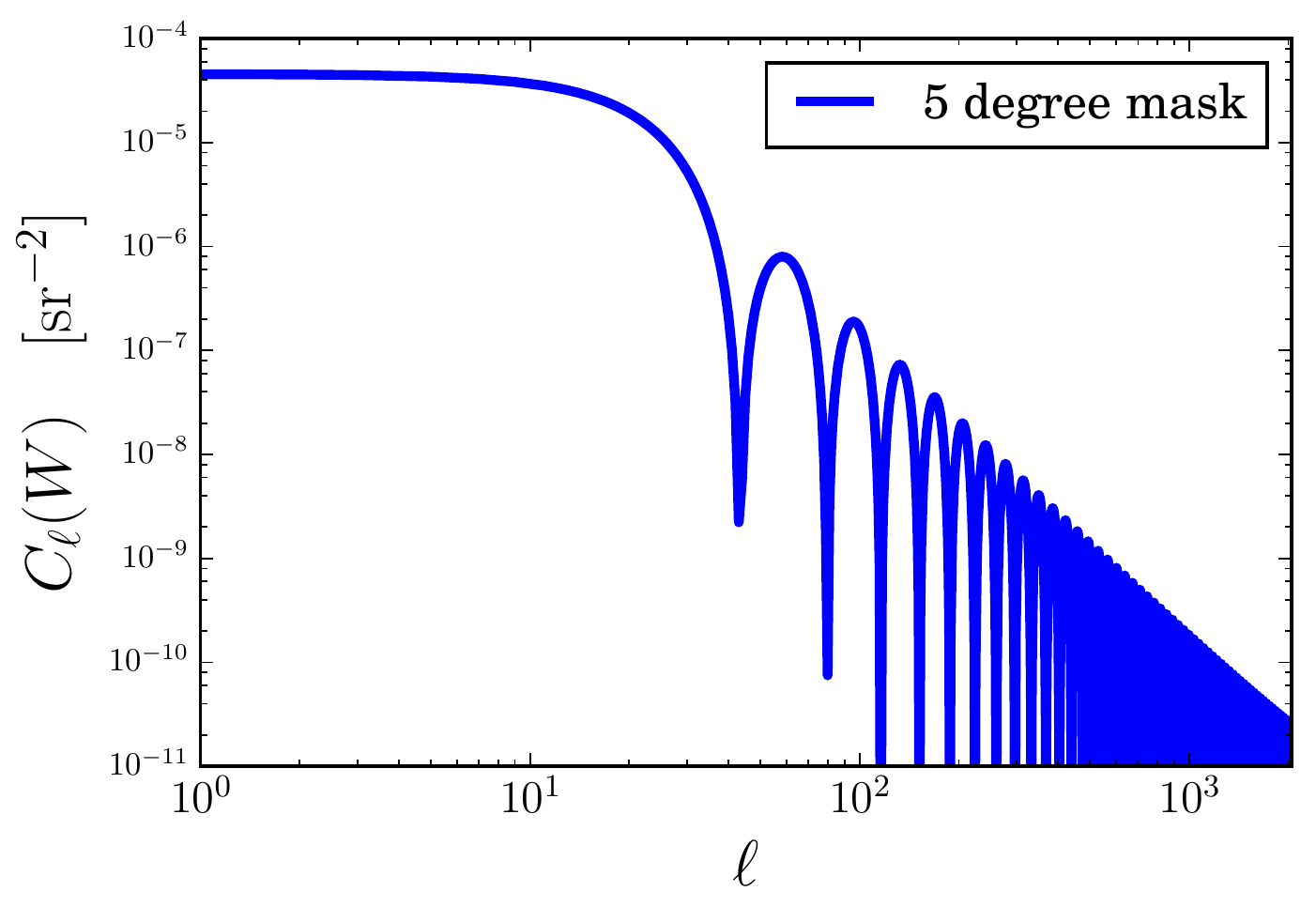}
\caption{Angular power spectrum of the 5 degree polar cap mask.}
\label{Fig:cl-mask5}
\end{center}
\end{figure}

Second, we can compare the results numerically for a mask with small enough sky coverage. To do so, we created a polar cap mask of radius 5 degree\footnote{As the power spectrum of a map is invariant under SO(3), centring the mask on the north pole is simply a convenient choice.}. The power spectrum of this mask is shown in Fig.~\ref{Fig:cl-mask5}.

Compared to Fig.~\ref{Fig:clmasks}, we see that power extends to smaller angular scales or higher multipoles. In this case, we found that we had to extend the sum in Eq.~(\ref{Eq:covSSC-sum-covell}) to higher multipoles, $\ell_\mr{max}=250$ to reach 1\% level convergence of the covariance prediction. 
This stays however numerically tractable, through the observation that the $\ell$ dependence of $\Cov_\ell^\mr{SSC}$ is very smooth, especially after $\ell_\mr{peak}=25$. Thus we can sample this dependence only for a small number of logarithmically spaced multipoles and interpolate when computing the sum in Eq.~(\ref{Eq:covSSC-sum-covell}).

\begin{figure}[!th]
\begin{center}
\includegraphics[width=0.99\linewidth]{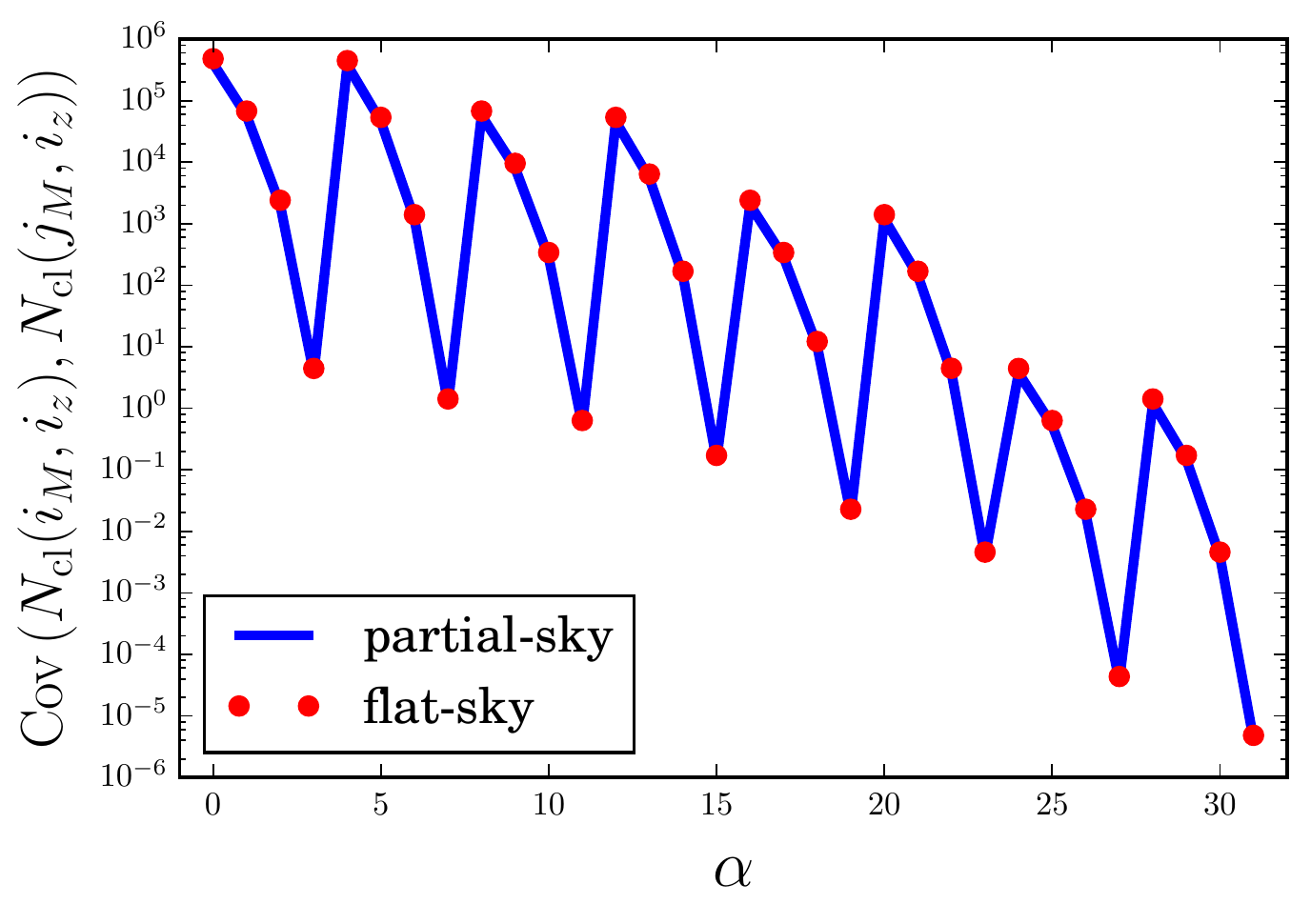}
\caption{Comparison of the SSC covariances derived from the partial sky formalism and the flat sky formula for a 5 deg circular sky patch. Auto-z covariances are shown as a function of $\alpha=j_M+n_M \, i_M+n_M^2 \, i_z$ (see text for details).}
\label{Fig:comp-flatsky-vs-psky}
\end{center}
\end{figure}

We found good agreement between the covariance derived from the partial sky formalism and that derived from the flat sky limit Eq.~(\ref{Eq:sigma2_flatsky}). This is visible, for example in Fig. \ref{Fig:comp-flatsky-vs-psky} showing both auto-z covariances as a function of $\alpha=j_M+n_M \, i_M+n_M^2 \, i_z$, i.e. the same ordering as in Fig.~\ref{Fig:ratio_Sij-KE_overflatsky}.

The partial sky formalism thus successfully recovers the flat sky approximation in the flat sky limit. Numerically, we even found the partial sky formalism to be faster than the full computation of Eq.~(\ref{Eq:sigma2_flatsky}), and only three times slower than the Sij approximation Eq.~(\ref{Eq:cov-SSC-Sij}).

%%%%%%%%%%%%%%%%%%%%%%%%%%%%%%%%%%%%%%%%%%%%%%%%%%%%%%%%%%%%%%%%%%%%
\section{Conclusion}\label{Sect:conclusion}

Super-sample covariance, also often called sample variance, is the dominant error for cluster counts at low cluster masses. For instance, \cite{Hu2006} have shown that even for a survey radius $\theta=2.4$ deg, the sample covariance is of the same order of magnitude as shot noise for cluster counts above $\log M=14.2$, although shot noise dominates for a threshold $\log M=14.4$. As shot-noise decreases faster with survey area than SSC, careful predictions of SSC become crucial for current and future surveys covering ever larger sky areas. It becomes even more crucial as these surveys are able to probe lower cluster masses through a higher density of galaxy detections.  Super-sample covariance is also crucial to the analysis of galaxy clustering and lensing shear, where it dominates statistical errors on small scales, and for probe combinations as it has been shown to couple probes \citep{TakadaBridle2007, Takada2014, Lacasa2016}.

In the case of cluster counts covariance, we examined theoretical SSC computation methods in the flat sky limit, comparing two analytical approximations proposed in the literature to a full computation. We found that both approximations underpredict the auto-$z$ covariance by 15\% to 30-35\%.

We then presented a harmonic expansion method for efficiently and accurately computing SSC for an arbitrary survey window function. We developed the method in the case of cluster counts, but it can be straightforwardly generalised to other probes such as galaxy clustering or lensing shear. Our derived expressions generalise previous full sky and flat sky equations found in the literature, properly reducing to these equations in the corresponding limits.
We have cast the final covariance expression from the partial sky formalism in a way that allows easy modification of the survey mask. This is particularly suitable for comparison of surveys, design of survey strategy, and tests of data cuts due to quality selection criteria or systematics.

When applying our partial sky formalism to a mask broadly similar to the DES footprint with $f_\mr{SKY}\sim 10\%$, we found a $\sim -15\%$ cross-$z$ covariance, meaning that the observables in the two redshift bins considered are anti-correlated. Hence the covariance matrix cannot be taken as block diagonal, as is the case for the approximation by \cite{Krause2016}, which is however restricted to the flat sky limit.
We also examined the possibility that the survey area is further reduced by pixel removal due to bright stars or systematics, for example. We found that this does not have important effects on the structure of the SSC matrix, only rescaling its amplitude by the effective survey area.

The results presented in this article thus render possible the theoretical computation of LSS covariances that account for selection and mask effect and also vary as a function of model parameters, as is the case in CMB analyses, for example. The latter parameter dependence is important in the case of likelihood analysis of cluster constraints, as it allows for self-calibration of the cluster observable-mass relation \citep{Lima2005,Hu2006,Baxter2016}. For general probes, it can also improve cosmological parameter constraints from likelihood inference analyses compared to methods that either neglect these effects or fix the covariance from data or simulations, thereby avoiding the risk of fixing the covariance at a potentially incorrect cosmology.

%%%%%%%%%%%%%%%%%%%%%%%%%%%%
\section*{Acknowledgements}
\vspace{0.2cm}

We thank Flavia Sobreira for providing us with a mask of a DES-like survey footprint.
We acknowledge the use of the {\tt Healpix} package by \cite{Gorski2005}. FL acknowledges support by the Swiss National Science Foundation.
ML is partially supported by FAPESP and CNPq.
MA is supported by FAPESP and by CNPq grant number 165049/2017-0.

%%%%%%%%%%%%%%%%%%%%%%%%%%%%%%%%%%%%%%%%%%%%%%%%%%%%%%%%%%%%%%%%%%%%
\bibliographystyle{aa}
\bibliography{bibliography}
%%%%%%%%%%%%%%%%%%%%%%%%%%%%%%%%%%%%%%%%%%%%%%%%%%%%%%%%%%%%%%%%%%%%

\appendix

\section{Redshift-dependent mask}\label{App:maskz}
In the case where the survey angular mask $ W$ depends on redshift (e.g. when there are significant depth variations in the sky), the definition of the background mode Eq.~(\ref{Eq:def-deltab}) is simply changed to
\be
\delta_b(z) = \frac{1}{\Omega_S(z)} \int \dd^2 \hn \ W(\hn,z) \ \delta(r\hn,z)\,,
\ee
where now the survey solid angle $\Omega_S$  depends on redshift.

Then Eq.~(\ref{Eq:sigma2-psky}) for $\sigma^2(z_1,z_2)$ is changed to
\ba\label{Eq:sigma2-zdependentmask}
\sigma^2(z_1,z_2) = \frac{1}{\Omega_S(z_1) \, \Omega_S(z_2)} \sum_\ell &\left\{ \right.(2\ell+1) \ C_\ell[W(z_1),W(z_2)] \ \nonumber \\ &\times C_\ell^m(z_1,z_2) \left.\right\}\,,
\ea
where $C_\ell[W(z_1),W(z_2)]$ is the angular cross-spectrum between the two masks at the two redshifts. This equation can then be integrated over $(z_1,z_2)$ through Eq.~(\ref{Eq:SSC-general}) to yield the SSC covariance.

We note that in this redshift-dependent case, we can no longer permute the $k$ and $z$ integrals as was done in Sect.~\ref{Sect:implementation}, which yielded a numerically efficient method.
Recent advances by \cite{Campagne2017} may however render Eq.~\ref{Eq:sigma2-zdependentmask} numerically computable through the AngPow software.

%%%%%%%%%%%%%%%%%%%%%%%%%%%%%%%%%%%%%%%%%%%%%%%%%%%%%%%%%%%%%%%%%%%%

\section{SSC for other probes}\label{App:other-probes}
The main results of this article were derived for the covariance of cluster number counts. However, it is straightforward to generalise the equations to other LSS probes, such as galaxy clustering and lensing shear. In those cases, the covariance equations are simpler for the angular power spectrum $C_\ell$ than for the angular correlation function. For instance, using Limber's approximation, \cite{Lacasa2016} have shown that the SSC of the galaxy angular power spectrum is given by
\ba
\nonumber \Cov\left(C_{\ell_1}^\mr{gal}(i_z),C_{\ell_2}^\mr{gal}(j_z)\right)& =  \int \dd V_{12} \ \frac{\nbargal(z_1) \, \nbargal(z_2)}{\Delta N_\mr{gal}(i_z) \, \Delta N_\mr{gal}(j_z)} \\
\nonumber & \times \frac{\partial P_\mr{gal}(k_{\ell_1},z_1)}{\partial \delta_b} \ \frac{\partial P_\mr{gal}(k_{\ell_2},z_2)}{\partial \delta_b} \\
& \times \sigma^2(z_1,z_2) \,,
\ea
where $k_{\ell_i}=(\ell_i+1/2)/r(z_i)$.

For lensing shear, we would get a similar equation, replacing the galaxy number density by the lensing selection function and the 3D galaxy power spectrum by the matter power spectrum. The equation can thus be generalised to
\ba
\nonumber \Cov\left(C_{\ell_1}^\alpha(i_z),C_{\ell_2}^\alpha(j_z)\right) & =  \int \dd V_{12} \ W_\alpha(z_1) \, W_\alpha(z_2) \\
\nonumber & \times \frac{\partial P_\alpha(k_{\ell_1},z_1)}{\partial \delta_b} \ \frac{\partial P_\alpha(k_{\ell_2},z_2)}{\partial \delta_b} \\
& \times \sigma^2(z_1,z_2)\,,
\ea
where the $\alpha$ index refers to either lensing or galaxy, and $W_\alpha$ is the corresponding selection function.
In that case, it is straightforward to generalise Eq.~(\ref{Eq:covellSSC-intk}), even including the possibility of the cross-covariance between galaxy and shear, \ba
\nonumber \Cov_\ell\left(C_{\ell_1}^\alpha(i_z),C_{\ell_2}^\beta(j_z)\right) &= \frac{1}{2\pi^2} \int k^2 \, \dd k \ P_m(k|z=0)\\
& \quad \times \Psi_\ell^{\ell_1,\alpha}(k|i_z) \ \Psi_\ell^{\ell_2,\beta}(k|j_z)\,,
\ea
where
\be
\Psi_\ell^{\ell_i,\alpha}(k|i_z) \equiv \int_{z\in \mr{bin}(i_z)} \dd V \, G(z) \, W_\alpha(z) \, \frac{\partial P_\alpha(k_{\ell_i},z)}{\partial \delta_b} \, j_\ell(k r)\,,
\ee

and the cross-covariance between cluster counts and either galaxy or shear,
\ba
\nonumber \Cov_\ell\left(\Ncl(i_M,i_z),C_{\ell_2}^\beta(j_z)\right) &= \frac{1}{2\pi^2} \int k^2 \, \dd k \ P_m(k|z=0)\\
& \quad \times \Psi_\ell^{n_h}(k|i_M,i_z) \, \Psi_\ell^{\ell_2,\beta}(k|j_z)\,,
\ea
which gives us all the equations needed to compute the auto and cross-covariances of cluster counts, galaxy angular power spectrum, and lensing shear power spectrum, i.e. the three main cosmological probes of current and future photometric galaxy surveys.

%%%%%%%%%%%%%%%%%%%%%%%%%%%%%%%%%%%%%%%%%%%%%%%%%%%%%%%%%%%%%%%%%%%%

\section{First multipoles}\label{App:first-multipoles}

The spherical Bessel functions $j_n(x)$ obey the recurrence relation
\be\label{Eq:jn-recurs}
j_{n+1}(x) = \frac{2n+1}{x} j_n(x) - j_{n-1}(x)\,,
\ee
such that they can be written analytically in terms of sines, cosines and polynomials, given the initial conditions
\ba
j_{-1}(x) &= \frac{\cos(x)}{x}\,,\\
j_0(x) &= \frac{\sin(x)}{x}\,.
\ea

The first few spherical Bessel functions are given by
\ba
\label{Eq:j0} j_0(x) &= \frac{\sin(x)}{x}\,,\\
\label{Eq:j1} j_1(x) &= \frac{1}{x}\frac{\sin(x)}{x} - \frac{\cos(x)}{x}\,,\\
\label{Eq:j2} j_2(x) &= \left(\frac{3}{x^2}-1\right)\frac{\sin(x)}{x} - \frac{3}{x}\frac{\cos(x)}{x} \, . 
\ea

We are trying to compute the following integrals:
\be
C_\ell^\mr{m}(z_{12})= \frac{2}{\pi} \int k^2 \dd k \ j_\ell(k r_1) \, j_\ell(k r_2) \ P_m(k|z_{12})\,.
\ee

For $j_\ell(x)$ of the form $A_\ell(x)\sin(x)/x-B_\ell(x)\cos(x)/x$, this yields
\ba\label{Eq:Cl_firstmultipoles_AB}
\nonumber C_\ell^\mr{m}(z_{12}) &= \frac{2}{\pi} \int k^2 \dd k \ \frac{P_m(k|z_{12})}{k^2 r_1 r_2}\\
\nonumber &\times \frac{1}{2}\ \Big[\big(A_\ell(kr_1)A_\ell(kr_2)+B_\ell(kr_1)B_\ell(kr_2)\big)\\
\nonumber & \qquad \qquad \times \cos\big(k(r_1-r_2)\big)\\
\nonumber &+\big(-A_\ell(kr_1)A_\ell(kr_2)+B_\ell(kr_1)B_\ell(kr_2)\big) \\
\nonumber & \qquad \qquad \times \cos\big(k(r_1+r_2)\big)\\
\nonumber & -\big(A_\ell(kr_1)B_\ell(kr_2)-B_\ell(kr_1)A_\ell(kr_2)\big) \\
\nonumber & \qquad \qquad \times \sin\big(k(r_1-r_2)\big)\\
\nonumber & -\big(A_\ell(kr_1)B_\ell(kr_2)+B_\ell(kr_1)A_\ell(kr_2)\big) \\
& \qquad \qquad \times \sin\big(k(r_1+r_2)\big)\Big] \,,
\ea
where the four integrals are Fourier (sine or cosine) transforms and can be computed numerically with FFTs. Let us note
\ba
\nonumber I_c^+\big(f(k,r_1,r_2)\big) &\equiv \frac{2}{\pi} \int k^2 \dd k \ \frac{P_m(k|z_{12})}{k^2 r_1 r_2} \ f(k,r_1,r_2)  \\
& \qquad \qquad \times \cos\big(k(r_1+r_2)\big)\,, \\
\nonumber I_c^-\big(f(k,r_1,r_2)\big) &\equiv \frac{2}{\pi} \int k^2 \dd k \ \frac{P_m(k|z_{12})}{k^2 r_1 r_2} \ f(k,r_1,r_2) \\
& \qquad \qquad \times \cos\big(k(r_1-r_2)\big)\,,  \\
\nonumber I_s^+\big(f(k,r_1,r_2)\big) &\equiv \frac{2}{\pi} \int k^2 \dd k \ \frac{P_m(k|z_{12})}{k^2 r_1 r_2} \ f(k,r_1,r_2) \\
& \qquad \qquad \times \sin\big(k(r_1+r_2)\big)\,,  \\
\nonumber I_s^-\big(f(k,r_1,r_2)\big) &\equiv \frac{2}{\pi} \int k^2 \dd k \ \frac{P_m(k|z_{12})}{k^2 r_1 r_2} \ f(k,r_1,r_2) \\
& \qquad \qquad \times \sin\big(k(r_1-r_2)\big)\,.
\ea

Then Eq.~(\ref{Eq:Cl_firstmultipoles_AB}) can be rewritten as
\ba
\nonumber C_\ell^\mr{m}(z_{12}) &= \frac{1}{2}\Big[
I_c^-\big(A_{\ell,1}A_{\ell,2}+B_{\ell,1}B_{\ell,2}\big)\\
\nonumber & \qquad +I_c^+\big(-A_{\ell,1}A_{\ell,2}+B_{\ell,1}B_{\ell,2}\big)\\
\nonumber & \qquad -I_s^-\big(A_{\ell,1}B_{\ell,2}-B_{\ell,1}A_{\ell,2}\big)\\
& \qquad -I_s^+\big(A_{\ell,1}B_{\ell,2}+B_{\ell,1}A_{\ell,2}\big)
\Big]\,,
\ea
with $A_{\ell,i}=A_\ell(k r_i)$.
For $\ell=0$ we have $A_0=1$ and $B_0=0$, thus
\be
C_0^\mr{m}(z_{12}) = \frac{1}{2}\ \Big[
I_c^-\big(1\big)-I_c^+\big(1\big)\Big]\,.
\ee

For $\ell=1$ we have $A_1=1/x$ and $B_1=1$. Therefore 
\ba
\nonumber C_1^\mr{m}(z_{12}) = \frac{1}{2} 
\Bigg[
&I_c^-\left(\frac{1}{k^2 r_1 r_2}+1\right)
+I_c^+\left(-\frac{1}{k^2 r_1 r_2}+1\right)  \nonumber \\
&-I_s^-\left(\frac{1}{k r_1}-\frac{1}{k r_2}\right)
-I_s^+\left(\frac{1}{k r_1}+\frac{1}{k r_2}\right) 
\Bigg]\,.
\ea

If we define $I_{c/s,n}^{+/-} \equiv I_{c/s}^{+/-}\left(1/k^n\right)$, the above equations yield
\ba
C_0^\mr{m}(z_{12})&=\frac{1}{2}\ 
\Big[I_{c,0}^- - I_{c,0}^+\Big]\,,\\
\nonumber C_1^\mr{m}(z_{12})&=\frac{1}{2}\ 
\Bigg[I_{c,0}^- + I_{c,0}^+ 
-I_{s,1}^-\left(\frac{1}{r_1}-\frac{1}{r_2}\right) \\
& \qquad -I_{s,1}^+\left(\frac{1}{r_1}+\frac{1}{r_2}\right) + \frac{I_{c,2}^- - I_{c,2}^+}{r_1r_2}
\Bigg]\,,
\ea

For $\ell=2$ we have $A_2=3/x^2-1$ and $B_2=3/x$, thus
\ba
\nonumber C_2^\mr{m}(z_{12}) &= \frac{1}{2}\ \Bigg[
I_c^-\left(\frac{9}{k^4 r_1^2 r_2^2}+\frac{3}{k^2}\left(\frac{3}{r_1 r_2}-\frac{1}{r_1^2}-\frac{1}{r_2^2}\right)+1 \right)\\
\nonumber &+I_c^+\left(-\frac{9}{k^4 r_1^2 r_2^2}+\frac{3}{k^2}\left(\frac{3}{r_1 r_2}+\frac{1}{r_1^2}+\frac{1}{r_2^2}\right)-1\right)\\
\nonumber &-I_s^-\left(\frac{9}{k^3 r_1 r_2}+\frac{3}{k}\right)\times\left(\frac{1}{r_1}-\frac{1}{r_2}\right)\\
&-I_s^+\left(\frac{9}{k^3 r_1 r_2}-\frac{3}{k}\right)\times\left(\frac{1}{r_1}+\frac{1}{r_2}\right) 
\Bigg]\\
\nonumber &=\frac{1}{2}\ \Bigg[I_{c,0}^- - I_{c,0}^+
-  3 \, I_{s,1}^-\left(\frac{1}{r_1}-\frac{1}{r_2}\right)\\
\nonumber &+3  \, I_{s,1}^+\left(\frac{1}{r_1}+\frac{1}{r_2}\right)
+3\,I_{c,2}^-\left(\frac{3}{r_1 r_2}-\frac{1}{r_1^2}-\frac{1}{r_2^2}\right)\\
\nonumber &+3\,I_{c,2}^+\left(\frac{3}{r_1 r_2}+\frac{1}{r_1^2}+\frac{1}{r_2^2}\right)
-\frac{9 \, I_{s,3}^-}{r_1 r_2}\left(\frac{1}{r_1}-\frac{1}{r_2}\right)\\&-\frac{9 \, I_{s,3}^+}{r_1 r_2}\left(\frac{1}{r_1}+\frac{1}{r_2}\right)
 +\frac{9\left(I_{c,4}^- - I_{c,4}^+\right)}{r_1^2 r_2^2}\Bigg]\,. 
 \label{Eq:C2-analyt}
\ea 

We note that $A_\ell$ and $B_\ell$ follow the same recurrence relation Eq.~(\ref{Eq:jn-recurs}) as the spherical Bessel functions $j_n$. As such we can look at whether there is a recurrence relation that would allow us to get the analytical formula for a general $C_\ell^\mr{m}(z_{12})$. Our efforts in this direction have shown only partially fruitful and are described below. 
We have
\ba
\nonumber C_{\ell+1}^\mr{m}(z_{12}) &= \frac{1}{2}\Big[
I_c^-\big(A_{{\ell+1},1}A_{{\ell+1},2}+B_{{\ell+1},1}B_{{\ell+1},2}\big) \nonumber \\
&\quad +I_c^+\big(-A_{{\ell+1},1}A_{{\ell+1},2}+B_{{\ell+1},1}B_{{\ell+1},2}\big)\nonumber \\
&\quad -I_s^-\big(A_{{\ell+1},1}B_{{\ell+1},2}-B_{{\ell+1},1}A_{{\ell+1},2}\big) \nonumber \\
&\quad -I_s^+\big(A_{{\ell+1},1}B_{{\ell+1},2} +B_{{\ell+1},1}A_{{\ell+1},2}\big)
\Big] \nonumber \\
&\equiv \frac{1}{2} \left(C_{\ell+1}^{c,-}+C_{\ell+1}^{c,+}-C_{\ell+1}^{s,-}-C_{\ell+1}^{s,+}\right)\,,
\ea
and then
\ba
\nonumber C_{\ell+1}^{c,-} =& I_c^-\bigg[
\left(\frac{2\ell+1}{k r_1}A_{\ell,1}-A_{\ell-1,1}\right)\left(\frac{2\ell+1}{k r_2}A_{\ell,2}-A_{\ell-1,2}\right)\\
&+
\left(\frac{2\ell+1}{k r_1}B_{\ell,1}-B_{\ell-1,1}\right)\left(\frac{2\ell+1}{k r_2}B_{\ell,2}-B_{\ell-1,2}\right)
\bigg]\\
\nonumber=& \ C_{\ell-1}^{c,-}+\frac{(2\ell+1)^2}{r_1 r_2} I_c^-\big[\left(A_{{\ell},1}A_{{\ell},2}+B_{{\ell},1}B_{{\ell},2}\right)/k^2\big]\\
\nonumber &-(2\ell+1) \ I_c^-\big[(A_{{\ell-1},1}A_{{\ell},2}+B_{{\ell-1},1}B_{{\ell},2})/kr_1\\
&\qquad\qquad +(1\leftrightarrow2)\big] \label{Eq:C_{l+1}^{c,-}}\,.
\ea

Now we define the following quantities
\ba
D_\ell^{c,-} &\equiv I_c^-\big[(A_{{\ell-1},1}A_{{\ell},2}+B_{{\ell-1},1}B_{{\ell},2})/kr_1+(1\leftrightarrow2)\big] \,,\\
E_\ell^{c,-} &\equiv I_c^-\big[(A_{{\ell-1},1}A_{{\ell},2}+B_{{\ell-1},1}B_{{\ell},2})/kr_2+(1\leftrightarrow2)\big] \,,
\ea
and decompose $C_\ell^{c,-}$, $D_\ell^{c,-}$ and $E_\ell^{c,-}$ onto the basis $I_{c,i}^-$
\ba
C_\ell^{c,-} &= \sum_{i=0}^{+\infty} \alpha^{c,-}_{\ell,i}(r_1,r_2) I_{c,i}^- \,,\\
D_\ell^{c,-} &= \sum_{i=0}^{+\infty} \beta^{c,-}_{\ell,i}(r_1,r_2) I_{c,i}^- \,,\\
E_\ell^{c,-} &= \sum_{i=0}^{+\infty} \gamma^{c,-}_{\ell,i}(r_1,r_2) I_{c,i}^- \,.
\ea

Then Eq.~(\ref{Eq:C_{l+1}^{c,-}}) gives a first recurrence relation
\ba\label{Eq:recur-rel-alphac-}
\alpha^{c,-}_{\ell+1,i} = \alpha^{c,-}_{\ell-1,i} + \frac{(2\ell+1)^2}{r_1 r_2}\alpha^{c,-}_{\ell,i-2}-(2\ell+1)\beta^{c,-}_{\ell,i}\,.
\ea

By computing $D_{\ell+1}^{c,-}$ and $E_{\ell+1}^{c,-}$, we can see that we have the two other recurrence relations~
\ba
\label{Eq:recur-rel-betac-} \beta^{c,-}_{\ell+1,i} &= (2\ell+1)\frac{2}{r_1 r_2}\alpha^{c,-}_{\ell,i-2}-\gamma^{c,-}_{\ell,i} \,, \\
\label{Eq:recur-rel-gammac-} \gamma^{c,-}_{\ell+1,i} &= (2\ell+1)\left(\frac{1}{r_1^2}+\frac{1}{r_2^2}\right)\alpha^{c,-}_{\ell,i-2}-\beta^{c,-}_{\ell,i} \,.
\ea

The system is closed through the initial conditions
\ba
\alpha^{c,-}_{0,0}&=1  &\alpha^{c,-}_{0,2}&=0  &\alpha^{c,-}_{0,i}&=0 \ \mr{for} \ i\geq3\\
\alpha^{c,-}_{1,0}&=1 &\alpha^{c,-}_{1,2}&=\frac{1}{r_1 r_2}  &\alpha^{c,-}_{1,i}&=0 \ \mr{for} \ i\geq3\\
\beta^{c,-}_{1,0} &= 0 &\beta^{c,-}_{1,2} &= \frac{2}{r_1 r_2} &\beta^{c,-}_{1,i}&=0 \ \mr{for} \ i\geq3\\
\gamma^{c,-}_{1,0} &=0 &\gamma^{c,-}_{1,2} &= \left(\frac{1}{r_1^2}+\frac{1}{r_2^2}\right) &\gamma^{c,-}_{1,i}&=0 \ \mr{for} \ i\geq3
\ea

The following properties can easily be shown by recurrence:
\ba
\forall \ell \qquad \alpha^{c,-}_{\ell,0}&=1\,, \qquad \beta^{c,-}_{\ell,0}=0\,, \qquad \gamma^{c,-}_{\ell,0}=0\,.\\
\forall \ell \qquad \alpha^{c,-}_{\ell,i}&=0\,, \qquad \mr{if} \ i \ \mr{odd\ or}\ i>2\ell\,. \\
\forall \ell \qquad \alpha^{c,-}_{\ell,2\ell}&=\frac{\prod_{n=0}^{\ell-1} (2n+1)^2}{(r_1 r_2)^\ell}=\frac{(2\ell-1)!^2}{(4 r_1 r_2)^\ell \ \ell!^2} \,.
\ea

With a bit more work, we can in principle solve for $\beta^{c,-}_{\ell,2}$ and $\gamma^{c,-}_{\ell,2}$; since $\alpha^{c,-}_{\ell,0}=1$,
\ba
\beta^{c,-}_{\ell+1,2} &= (2\ell+1)\frac{2}{r_1 r_2}-\gamma^{c,-}_{\ell,2} \,, \\
\gamma^{c,-}_{\ell+1,2} &= (2\ell+1)\left(\frac{1}{r_1^2}+\frac{1}{r_2^2}\right)-\beta^{c,-}_{\ell,2}\,.
\ea

This can be put in the form
\ba
X_{\ell+1} &= M X_\ell + V_\ell\,,
\label{Eq:Xell}
\ea
for
\ba
X_\ell&=\begin{pmatrix}\beta^{c,-}_{\ell,2} \\ \gamma^{c,-}_{\ell,2}\end{pmatrix}\,, \\ 
M&=\begin{pmatrix} 0 & -1 \\ -1 & 0\end{pmatrix} \,,\\ 
V_\ell&=(2\ell+1)\begin{pmatrix} \frac{2}{r_1 r_2} \\ \frac{1}{r_1^2}+\frac{1}{r_2^2}\end{pmatrix}=(2\ell+1)X_1\,.
\ea

Finally, Eq.~(\ref{Eq:Xell}) has the following solution:
\be
X_\ell = \sum_{k=0}^{\ell-1} M^{\ell-k-1} \ (2k+1)X_1 \,,
\ee
which in principle gives the solution for $\beta^{c,-}_{\ell,2}$ and $\gamma^{c,-}_{\ell,2}$. We can insert this into Eq.~(\ref{Eq:recur-rel-alphac-}) for $i=2$ to get a closed recurrence relation for $\alpha_{\ell,2}^{c,-}$ and solve for it. In turn, this can be inserted into Eqs.~(\ref{Eq:recur-rel-betac-}) and (\ref{Eq:recur-rel-gammac-}) for $i=4$ to get a closed recurrence relation for $(\beta_{\ell,4}^{c,-},\gamma_{\ell,4}^{c,-})$, and so on. So this represents a procedure for solving this set of equations and get analytical expressions for $C_\ell^{c,-}$. Similarly, we could derive recurrence relations for $C_\ell^{c,+}$, $C_\ell^{s,-}$, and $C_\ell^{s,+}$. In practice this represents a daunting task, as even the formula for $\ell=2$ (Eq.~(\ref{Eq:C2-analyt})) is already cumbersome.

Moreover, we may expect that these analytical formulae are doomed to present numerical instability at high $\ell$. Indeed, the expansion of the spherical Bessel functions in terms of powers of $1/x$ as in Eqs.~(\ref{Eq:j0})-(\ref{Eq:j1})-(\ref{Eq:j2}) is ill-advised at high $\ell$, as it leads to delicate cancellations of the numerous terms, in particular for $x\leq\ell$. We thus expect that our analytical formulae for $C_\ell$ also leads to delicate cancellations that may be numerically unstable. 
Another way of seeing this is to notice that, at high $\ell$, we have from Limber's approximation that $C_\ell\propto P_m\left[k=(\ell+1/2)/r(z)\right]$, and is thus a decreasing function of $\ell$. However we saw that $\alpha^{c,-}_{\ell,0}=1$ and thus $C_\ell$ always contains a term $I_{c,0}^-/2\approx C_0$. This term hence needs to be (at least partially) cancelled by higher order terms, and this cancellation needs to be increasingly precise at high $\ell$, since $P_m(k)$ is a steep decreasing function of $k$.

Given these issues, in practice, we implemented this low multipole method only for $\ell=0,1,2$, and used these to check our results obtained from the numerical method described in Sect.~\ref{Sect:implementation}.

%%%%%%%%%%%%%%%%%%%%%%%%%%%%%%%%%%%%%%%%%%%%%%%%%%%%%%%%%%%%%%%%%%%%

\end{document}